\documentclass[hidelinks,12pt]{article}
\usepackage{hyperref,natbib,amsmath,amssymb, amsthm}
\usepackage{import, color,graphicx}

\usepackage{diagbox}

\newcommand{\blu}[1]{\textcolor{black}{#1}} 

\usepackage{epsf}
\usepackage{epsfig}
\usepackage{amsfonts}
\usepackage{setspace}
\usepackage{algpseudocode}
\usepackage{algorithm2e}
\usepackage[usenames, dvipsnames]{xcolor}

\usepackage{caption}
\usepackage{subcaption}
\usepackage[normalem]{ulem}

\newcommand{\blind}{0}

\newcommand{\mb}[1]{#1}

\newcommand{\vecX}{\mathbf{X}}
\newcommand{\vecY}{\mathbf{Y}}

\def\spacingset#1{\renewcommand{\baselinestretch}%
{#1}\small\normalsize} \spacingset{1}

\begin{document}

\if0\blind
{
\title{\vspace{-1cm}  Distance-distributed design\\ for Gaussian process surrogates}
\author{
	Boya Zhang\thanks{Corresponding author: Department of Statistics, Virginia Tech, \href{mailto:boya66@vt.edu}{\tt boya66@vt.edu}}\
\and 
D.~Austin Cole\thanks{Department of Statistics, Virginia Tech}
\and 
Robert B.~Gramacy\footnotemark[2]
}
\date{}
\maketitle
}\fi

\if1\blind
{
  \bigskip
  \bigskip
  \bigskip
  \begin{center}
    {\LARGE\bf Distance-distributed design\\ for Gaussian process surrogates}
\end{center}
  \medskip
  \bigskip
} \fi

\begin{abstract}
A common challenge in computer experiments and related fields is to
efficiently explore the input space using a small number of samples, i.e., the
experimental design problem. Much of the recent focus in the computer
experiment literature, where modeling is often via Gaussian process (GP)
surrogates, has been on space-filling designs, via maximin distance, Latin
hypercube, etc. However, it is easy to demonstrate empirically that such
designs disappoint when the model hyperparameterization is unknown, and
must be estimated from data observed at the chosen design sites.  This is true
even when the performance metric is prediction-based, or when the target of
interest is inherently or eventually sequential in nature, such as in blackbox
(Bayesian) optimization. Here we expose such inefficiencies, showing that in
many cases a purely random design is superior to higher-powered alternatives. We
then propose a family of new schemes by reverse engineering the qualities of
the random designs which give the best estimates of GP lengthscales.
Specifically, we study the distribution of pairwise distances between design
elements, and develop a numerical scheme to optimize those distances for a
given sample size and dimension. We illustrate how our distance-based designs,
and their hybrids with more conventional space-filling schemes, outperform in
both static (one-shot design) and sequential settings.
\end{abstract}

\if0\blind
{
\bigskip
\noindent {\bf Keywords:}
Computer experiment, emulator, experimental design, sequential
design, Bayesian optimization, kriging, lengthscale  }\fi


\section{Introduction}
\label{sec:intro}

Computer simulation experiments are widely used in the applied sciences to
simulate time-consuming or costly physical, biological, or social dynamics.
Depending on the dynamics being simulated, these experiments can themselves be
computationally demanding, limiting the number of runs that can be entertained.
Design and meta-modeling considerations have spawned a research area at the
intersection of spatial modeling, optimization, sensitivity analysis, and
calibration. \citet{santner:etal:2003} provide an excellent review.

Gaussian process (GP) surrogates, originally for interpolating data from
deterministic computer simulations \citep{Sacks:1989}, have percolated to
the top of the hierarchy for many meta-modeling purposes.  GP surrogates are
fundamentally the same {\em kriging} from the spatial statistics literature
\citep{math:1963}, but generally applied in higher dimensional (i.e., $>$ 2d)
settings.  They are preferred for their simple, partially analytic,
nonparametric structure.  GPs' out-of-sample predictive accuracy and coverage
properties are integral to diverse applications such as Bayesian optimization
\citep[BO][]{jones:schonlau:welch:1998}, calibration
\citep{kenn:ohag:2001,higdon2004combining}, and input sensitivity analysis
\citep{SaltEtAl2008}.  Although there are many variations on GP specification,
\citet{chen2016} nicely summarize how such nuances often have little impact in
practice.

On the other hand, \citeauthor{chen2016} cite experimental design as playing
an out-sized role.  Despite GPs' elevation to ``canonical'' status as
surrogates, there has not been quite the same degree of confluence in how to
design a computer experiment for the purpose of such modeling.  In part this
is simply a consequence of different goals emitting different criteria for
valuing, and thus selecting, inputs. An exception may be the general agreement
that it is sensible, if possible, to proceed sequentially, either one point at
a time or in batches.  An underlying theme for static (all-at-once) design, or
for seeding a sequential design, has been to seek space-fillingness, where the
selected inputs are spread out across the study space.  For a nice review,
see~\citet{Pronzato:2011}.

There are many ways in which a design might be considered space-filling.
Maximin-distance and minimax-distance design \citep{Johnson:1990} are two
common approaches based on geometric criteria. A maximin design attempts to
make the smallest distance between neighboring points as large as possible;
conversely, minimax  attempts to minimize the maximum distance.
\blu{ A common variation on maximin is $\phi_p$ \citep{Morris:1995},
$$
\phi_p = \left[ \sum_{k=1}^K J_k d_k^{-p} \right]^{1/p}.
$$
where $d_k$ is one of the $K$ unique pairwise distances in a design and
$J_k$ is the number of pairs at that distance.\footnote{In most applications,
$K = {n \choose 2}$ and all $J_k = 1$.}  Designs obtained by minimizing
$\phi_p$ are actually maximin for all $p$, i.e., the smallest distance
$\min_k d_k$ is maximized.  At $p \rightarrow \infty$ the equivalence is
immediate; $\phi_p$ designs for smaller $p$ have greater spread in smaller
distances ($d_{(-k)}$).}

Alternatively, one may desire a design that spreads points evenly across the
range of each individual input, i.e., where projections on each dimension are
still space-filling.  Maximin and minimax designs do not produce such an
effect; in fact, they can be pathologically bad in this regard. Latin
hypercube sampling \citep[LHS,][]{Mckay:1979} can guarantee this {\em
one-dimensional uniformity} property.  For a nice review of LHS and other
space-filling designs for computer experiments, see \citet{lin:tiang:2015}.

Space-filling designs intuitively work well when prediction accuracy is of
primary interest, \blu{ seeking cover everywhere you might want to predict.}  
However, it is easy to show [as we do in Section
\ref{sec:randomtobeta}] that space-filling designs are inefficient for
learning GP hyperparameters, discussed in further detail in Section
\ref{sec:setup}.  It turns out that a random uniform design is actually better
than maximin, \blu{$\phi_p$} and LHS in that setting, \blu{echoing a rule-of-thumb
from variogram estimation with lattice data in geostatistics \citep{zhao:2004}.}

Considering that GP predictive prowess depends upon hyperparameterization,
good prediction results must tacitly depend upon fortuitously chosen
hyperparameters. If good settings are indeed known, then {\em model-based
design} represents an attractive alternative to (model-free) space-filling
design. Example criteria include maximizing the entropy between prior and
posterior (maximum entropy design), minimizing the integrated mean-squared prediction error
\citep[IMSPE, ][Chapter 6]{santner:etal:2003}, and Fisher information
\citep{Zimmerman:2006}.  These lead to nice sequential extensions, when
alternating between design and learning stages.  However, such schemes can
suffer when initialized poorly. 
Seemly optimal choices of seed design or hyperparameter can lead to pathologically poor performance.

Here we propose a new class of designs that attempts to resolve that
chicken-or-egg problem.  GP correlation structures are typically built upon
scaled pairwise distance calculations, so we hypothesize that certain sets of
pairwise distances offer a more favorable basis for estimating those scales: 
so-called GP {\em lengthscale} hyperparameters. The spirit of our study is
similar to that of \citet{Morris:1991}, but we take a more empirical approach
and ultimately provide a message that is more upbeat. Quite simply, we observe
the empirical distribution of pairwise distances of random designs which are
better than space-filling ones for the purpose of lengthscale estimation. We
then parameterize those distributions within the $\mathrm{Beta}(\alpha,
\beta)$ family, and propose a numerical optimization scheme to tune
$(\alpha,\beta)$  to design size $n$ and input dimension $d$.  \blu{In this way, our
methodology can be seen as a more aggressive and constructive variation of 
\citeauthor{zhao:2004}'s study for variograms.}

\blu{Despite sacrificing}  positional space-fillingness for
relative distance-fillingness in order to target hyperparameter estimation, we
show that \blu{``betadist'' designs} still perform favorably in prediction
exercises. Inspired by \citet{Morris:1995}'s hybridization of LHS and maximin,
we propose hybridizing LHS with betadist designs to strike a balance between
space and distance-filling toward \blu{even} more accurate prediction.

The remainder of the paper is organized as follows. In Section \ref{sec:setup}
we review GP modeling and design details pertinent to our methodological
contribution.  Section \ref{sec:randomtobeta} demonstrates how space-filling
designs fall short in certain respects, and proposes distance-based remedies
based on reverse engineering qualities of the best random designs. Section
\ref{sec:lhsbeta} explores hybrids of these betadist designs with LHS.
Illustrative examples and empirical comparisons are provided
throughout.  Section \ref{sec:seq} provides a comprehensive empirical
validation in two disparate sequential design settings, where betadist, LHS
hybrids and comparators are used to build initial/seed designs.  We conclude
in Section \ref{sec:discuss} with a brief discussion.

\section{Setup and related work}
\label{sec:setup}

Here we review essentials as a means of framing our contributions,
establishing notation, and connecting to related work on design and modeling
for computer  experiments.

\subsection{Gaussian Process surrogates}

Let $f: \mathbb{R}^d \rightarrow \mathbb{R}$, denote an unknown function,
generically, but standing in specifically for a computationally expensive
computer model simulation.  There is interest in limiting the evaluation of
$f$, so one designs an experimental plan of runs with the aim of fitting a
meta-model, e.g., a Gaussian process (GP), which can be used as a surrogate in
lieu of future expensive evaluations.  Let $\vecX$ = $\{\mathbf{x}_1, \dots,
\mathbf{x}_n\}$ denote the chosen $d$-dimensional design, and let $\vecY$ =
$(y_1,...,y_n)^\top$ collect outputs $y_i = f(\mathbf{x}_i)$, for $i=1,
\dots,n$.  

Putting a GP prior on $f$ amounts to specifying that any finite
realization of $f$, e.g., our $n$ observations $\vecY$, has a multivariate
normal (MVN) distribution.  MVNs are uniquely specified by a mean vector and
covariance matrix.  It is common in the computer experiments literature to
take the mean to be zero, and to specify the covariance structure via scaled
inverse Euclidean distances.  For example, $\vecY \sim
\mathcal{N}_n(\mathbf{0}, \mathbf{K}_n)$, where $\mathbf{K}_{ij}$ follows
\begin{equation}
\mathbf{K}_{ij} = k_{\tau^2, \theta}(\mathbf{x}_i, \mathbf{x}_j) = \tau^2
\exp\left\{-\frac{||\mathbf{x}_i - \mathbf{x}_j||^2}{\theta}\right\}. \label{eq:K}
\end{equation}
Above, $\tau^2$ is an amplitude hyperparameter, and $\theta$ is the
lengthscale, determining the rate of decay of correlation as a function of
distance in the input space.  This choice of correlation structure is called
the isotropic Gaussian family. Although we assume this structure throughout for
simplicity, we see no reason why our proposed methodology (which emphasizes
design, not modeling) could not be extended to other correlation families, 
or to the stochastic ($f + \varepsilon$)
setting via additional hyperparameters.  

Fixing $\theta$ and $\tau^2$, the GP predictive equations at new inputs
$\mathbf{x}$, given the data $(\vecX,
\vecY)$, have a convenient closed form derived from simple MVN
conditioning identities.  The (posterior) predictive distribution for
$Y(\mathbf{x}) \mid \vecY$ is Gaussian with
\begin{align}
  \mbox{mean} && \mu(\mathbf{x}\mid \vecY) &= \mathbf{k}^\top(\mathbf{x})  \mathbf{K}_n^{-1}\vecY,
\label{eq:predgp} \\
\mbox{and variance} &&
 \sigma^2(\mathbf{x} \mid \vecY) &=
k_{\tau^2, \theta}(\mathbf{x}, \mathbf{x}) - \mathbf{k}^\top(\mathbf{x})\mathbf{K}_n^{-1} \mathbf{k}(\mathbf{x}), \nonumber
\end{align}
where $\mathbf{k}^\top(\mathbf{x})$ is the $n$-vector whose $i^{\mbox{\tiny th}}$
component is $k_{\tau^2,\theta}(\mathbf{x}, \mathbf{x}_i)$.

Unknown hyperparameters can be inferred by viewing $\vecY \sim
\mathcal{N}_n(\mathbf{0}, \mathbf{K}_n)$ as a likelihood and maximizing its
logarithm numerically, or via Bayesian posterior sampling.  In the former case
(MLE), we obtain $\hat{\tau}^2 = n^{-1}
\mathbf{Y}^\top \mathbf{K}_n^{-1} \mathbf{Y}$ in closed form, which may be
used to derive a profile/concentrated multivariate Student-$t$ likelihood for
$\theta$.  In the Bayesian setting, $\tau^2$ may analytically be integrated
out under an inverse-Gamma prior \cite[see, e.g.,][]{gramacy:apley:2015}.
Either way, numerical methods are required to learn appropriate lengthscale
settings $\hat{\theta}$.  Throughout, we use {\tt mleGP} from the {\tt laGP}
library \citep{laGP} for {\sf R} \citep{cran:R} via ``L-BFGS-B''
\citep{byrd:etal:1995} leveraging closed form derivatives.

\subsection{Thinking about designs for GPs}

The prediction equations (\ref{eq:predgp}) suggest a space-filling training
design for $\vecX$ since $\sigma^2(\mathbf{x})$, for testing $\mathbf{x}$, is
quadratically related to distances to nearby $\mathbf{x}_i$ locations through
$\mathbf{k}(\mathbf{x})$.  However that tacitly assumes the hyperparameters,
particularly the lengthscale $\theta$, are known.  Where is a good $\theta$
supposed to come from?   While we acknowledge that it is sometimes possible to
intuit reasonable values or ranges for $\theta$, based on knowledge of the
underlying dynamics being modeled, such cases are rare in practice, and
useless as a default {\em modus operandi}, e.g., in software.  Thus our
presumption is that $\theta$ must be learned from data, which requires a
design. Intuitively, a space-filling design is poor for such purposes since
its deliberate inability to furnish short distances biases inference toward
longer lengthscales.

Sequential design, iterating between design and learning, has been suggested
as a remedy.  Yet space-filling design is still common in initial stages. For
example, \citet{Tan:2013} writes ``minimax designs are intended to be initial
designs for computer experiments, which are almost always sequential in
nature''. While we agree with the spirit of that statement, we disagree that
spreading out the points is the best way to seed this process.  The reason is
that subsequent sequential selections are usually model-based, e.g., via
$\sigma^2(\mathbf{x})$, and thus hyperparameter-sensitive. \blu{Note that in
sequential application, both IMSPE and maximum entropy-based designs are about
predictive variance.  The former maximizes integrated variance; the latter
maximizes directly.} \blu{One must be careful not to introduce} a feedback
loop where sequential decisions reinforce bad hyperparameters.

Some say that way out of that vicious cycle is to utilize other geometric
rather than model-based criteria for sequential selection, e.g., with
cascading LHSs \citep{Lin:2010}.  However, if the design goal is not directly
prediction-based, such as in BO \citep{jones:schonlau:welch:1998}, that
approach is clearly inefficient.  Plus in the BO literature, regularity
conditions underlying the theory for convergence (to global optima) insist on
fixed hyperparameterization.  This is specifically to avoid pathological
settings arising from feedback between sequential acquisition and inference
calculations~\citep{bull:2011}.

Perhaps our main thesis is that initial design for hyperparameter learning is
paramount to obtaining robust (good) behavior in repeated application.  While
some space-filling designs are better than others in this context, we observe
that it is important to be filling in a different sense.  Inference for
hyperparameters via the likelihood involves pairwise inverse distances
$\mathbf{x}_i - \mathbf{x}_j$ through $\mathbf{K}_{ij}$.  Therefore, it could
help to be more filling in that dimension.  As we show in Section
\ref{sec:randomtobeta}, simple random uniform designs are actually better than
the typical maximin and LHS alternatives, sometimes substantially so.
Intuitively, this is because random designs lead to a less clumpy, more
unimodal, distribution of relative distances compared to maximin, for example.
[See Figure \ref{fig:distdens} and surrounding discussion.]  Based on the
outcome of that study, we speculated that having a uniform distribution of
such pairwise distances---as opposed to uniform in position---would fare even
better.

That intuition turned out to be incorrect.  However initial investigations
pointed to a promising class of alternatives, targeting a more refined choice
of desirable pairwise distance distributions. Although the strategy we propose
imminently is novel in the context of design and analysis of computer
simulation experiments, it is not without precedent in the spatial statistics
literature, where variogram-based inference is, historically, at least as
common as likelihood-based methods \citep[see,
e.g.,][]{Russo:1984,Cressie1985}.  Out of that literature came the
rule-of-thumb that at least thirty pairs of data points should populate
certain distance strata. \citet{Morris:1991} subsequently revised that number
upwards, accounting for spatial correlations which devalue information
provided by nearby pairs.

The spirit of our contribution is similar to these works, although we shall
make no recommendations about design size.  Suggestions along these lines in
the computer experiments literature, such as $n=10d$ \citep{Loeppky:2009},
have been met with mixed reviews---never mind that the nuance of arguments
behind that particular suggestion is often forgotten. Instead, presuming small
fixed (initial) design sizes, we target the search for coordinates with
desirable qualities for lengthscale estimation. Our first idea ignores
position information entirely, focusing expressly on pairwise distances.  We
later revise that perspective to hybridize with LHS and acknowledge that a
degree of space-fillingness may be desirable when the over-arching modeling
goal is oriented toward prediction.

\section{Better than random}
\label{sec:randomtobeta}

Consider the following simple experiment in the input space $[0,1]^d$, for
$d=2,3,4,5,6$, taken in turn.  For thirty equally spaced ``true'' lengthscales
$\theta^{(t)} \in (0.1, \sqrt{d}]^d$, for $t = 1, \dots, 30$ we generate
$i=1,\dots,1000$ designs $\vecX^{(t,i)}$ of size $n=2^{d+1}$ and simulate
$\mathbf{Y}^{(t,i)}
\sim \mathcal{N}(\mathbf{0}, \mathbf{K}_n)$.  Entries of $\mathbf{K}_n$ are
calculated as in (\ref{eq:K}) via the rows of $\mathbf{X}^{(t,i)}$ and
hyperparameters $\tau^2=1$ and $\theta^{(t)}$.  Several design criteria are
discussed shortly. For each $(t,i)$, MLEs $\hat{\theta}^{(t,i)}$ are calculated
from data $(\vecX^{(t,i)},
\mathbf{Y}^{(t,i)})$. Finally, we collect average squared discrepancies
between estimated and true lengthscales via $\mathrm{logMSE}_t =
\log \left\{ \sum_{i=1}^{1000} (\hat{\theta}^{(t,i)} - \theta^{(t)})^2 \right\}$.

\begin{figure}[ht!]
\centering
\includegraphics[scale=0.6]{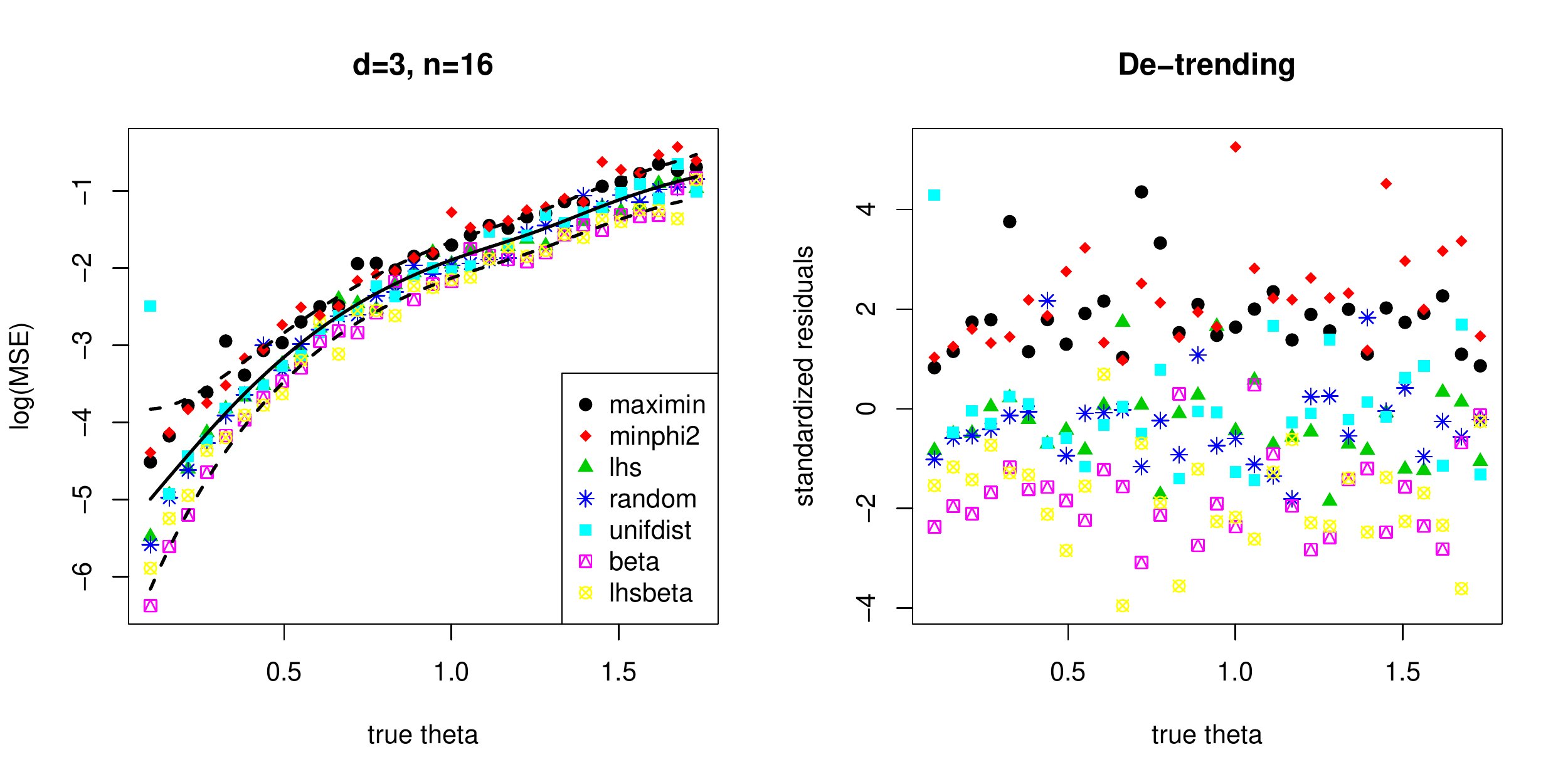}
\caption{logMSEs from design experiment and de-trending surface.}
\label{fig:rmses}
\end{figure}

As an example of the logMSEs obtained, the left panel of Figure
\ref{fig:rmses} shows the $(d=3, n=16)$ case.  The first thing to notice in
that plot is that as $\theta^{(t)}$ increases so does $\mathrm{logMSE}_t$, for
all design methods. Apparently, it is \blu{``harder''} to accurately estimate
lengthscales $\theta$ as they become longer. \blu{Harder is in quotes because
this metric} obscures the relative performance of the design methods, although
some \blu{consistently} stand out as worse (maximin/black circles) or better
(beta/pink squares or lhsbeta/yellow squares) than others.  To level the
playing field for subsequent analysis, we calculated standardized residuals
using a de-trending surface estimated from all of the dots, taken together.
To cope with the outliers we fit a heteroskedastic Student-$t$ GP as described
by \citet{Chung2018} and implemented in the {\tt hetGP} package \citep{hetGP}.
Section \ref{sec:betaopt} provides further details on our use of {\tt hetGP}
in this context.  Standardized residuals ($r_t =
(\mathrm{logMSE}_t-\mu_t)/\sigma_t$, with $\mu_t$ and $\sigma_t$ from {\tt
hetGP}), are shown in the right panel of the figure.

\begin{figure}[ht!]
	\begin{minipage}[b]{0.48\linewidth}
		\centering
		\includegraphics[width=\textwidth]{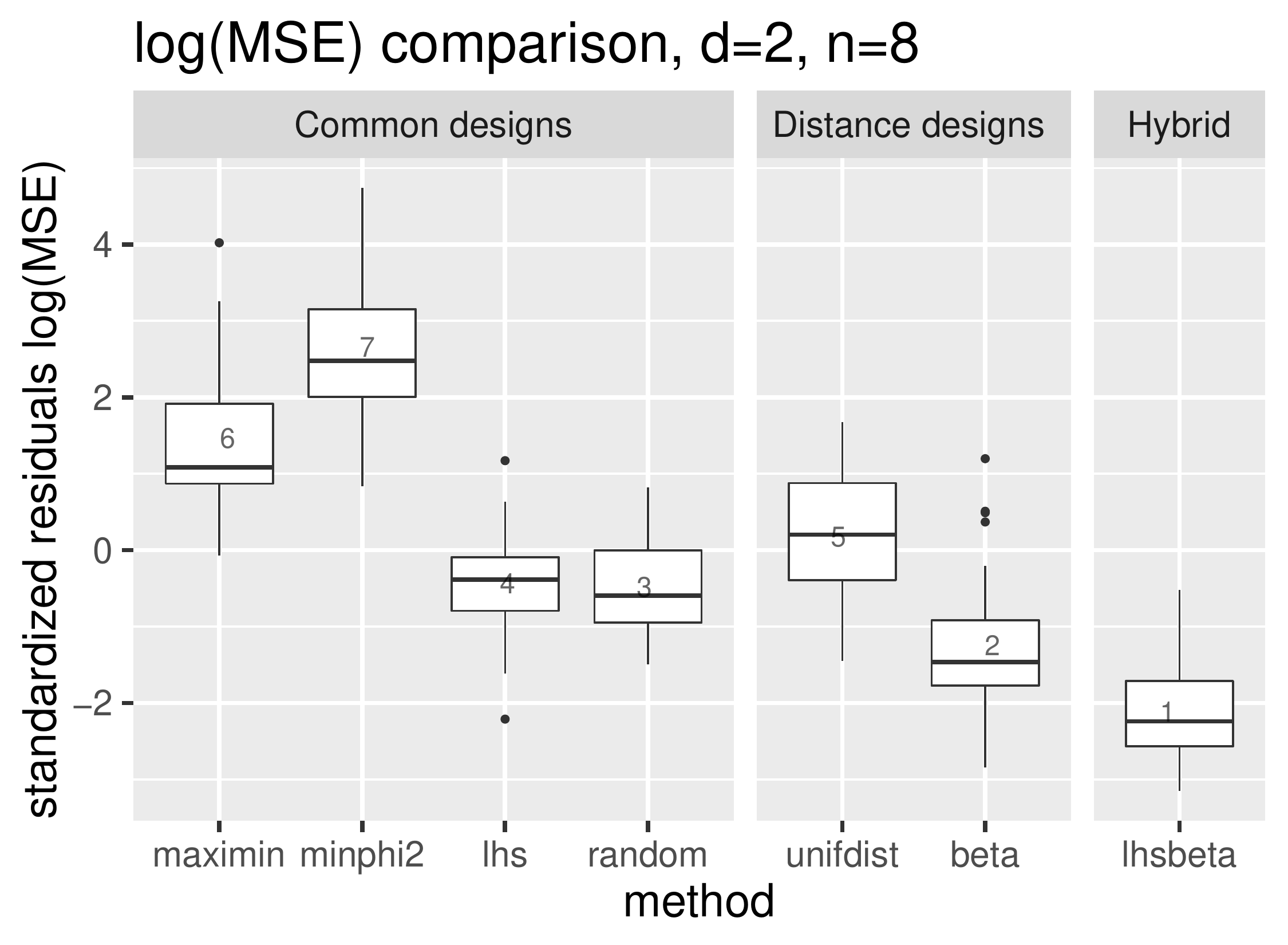}
	\end{minipage}%
\begin{minipage}[b]{0.48\linewidth}
	\centering
	\includegraphics[width=\textwidth]{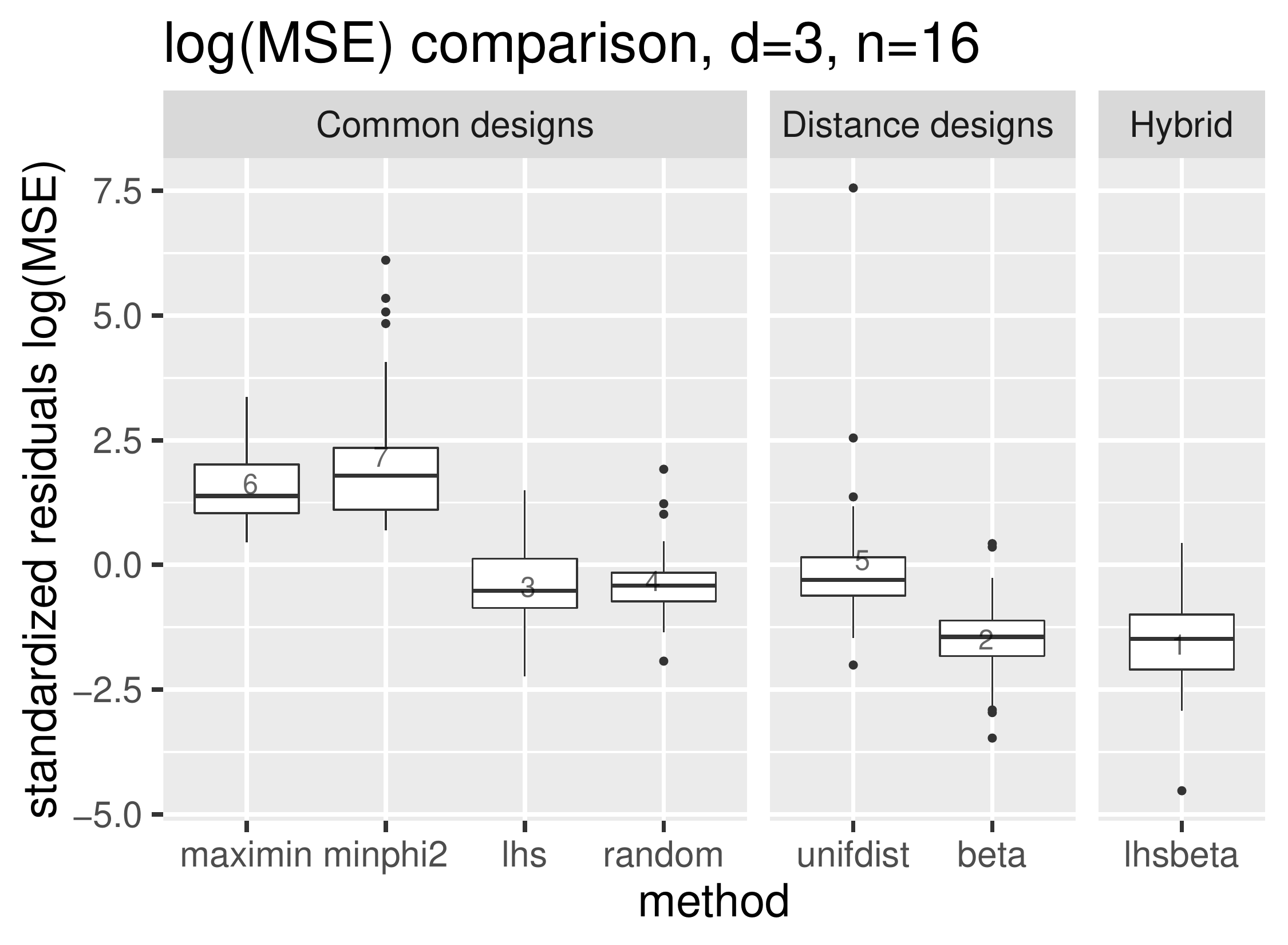}
\end{minipage}\quad

\begin{minipage}[b]{0.48\linewidth}
	\centering
	\includegraphics[width=\textwidth]{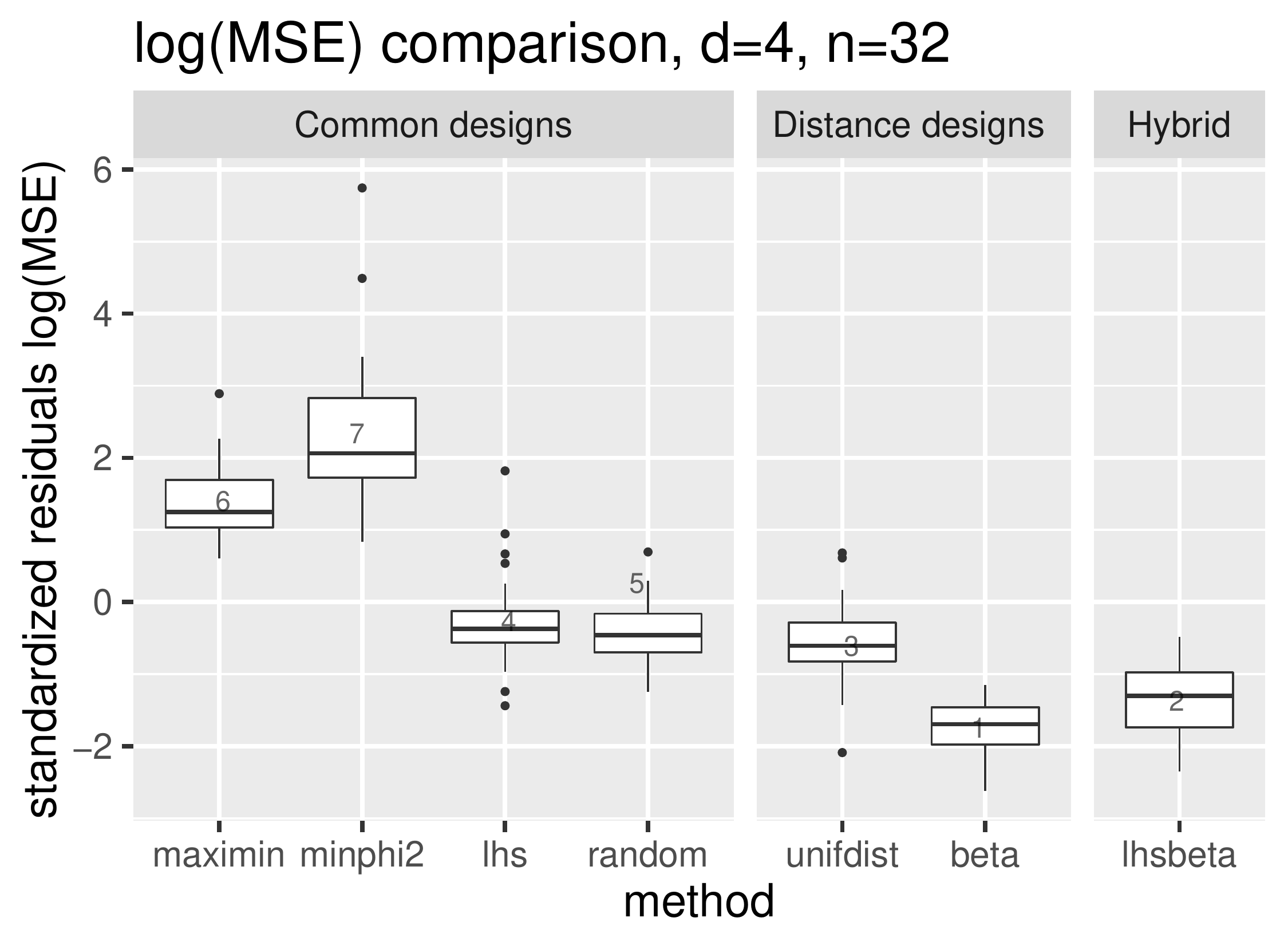}
\end{minipage}%
\begin{minipage}[b]{0.48\linewidth}
	\centering
	\includegraphics[width=\textwidth]{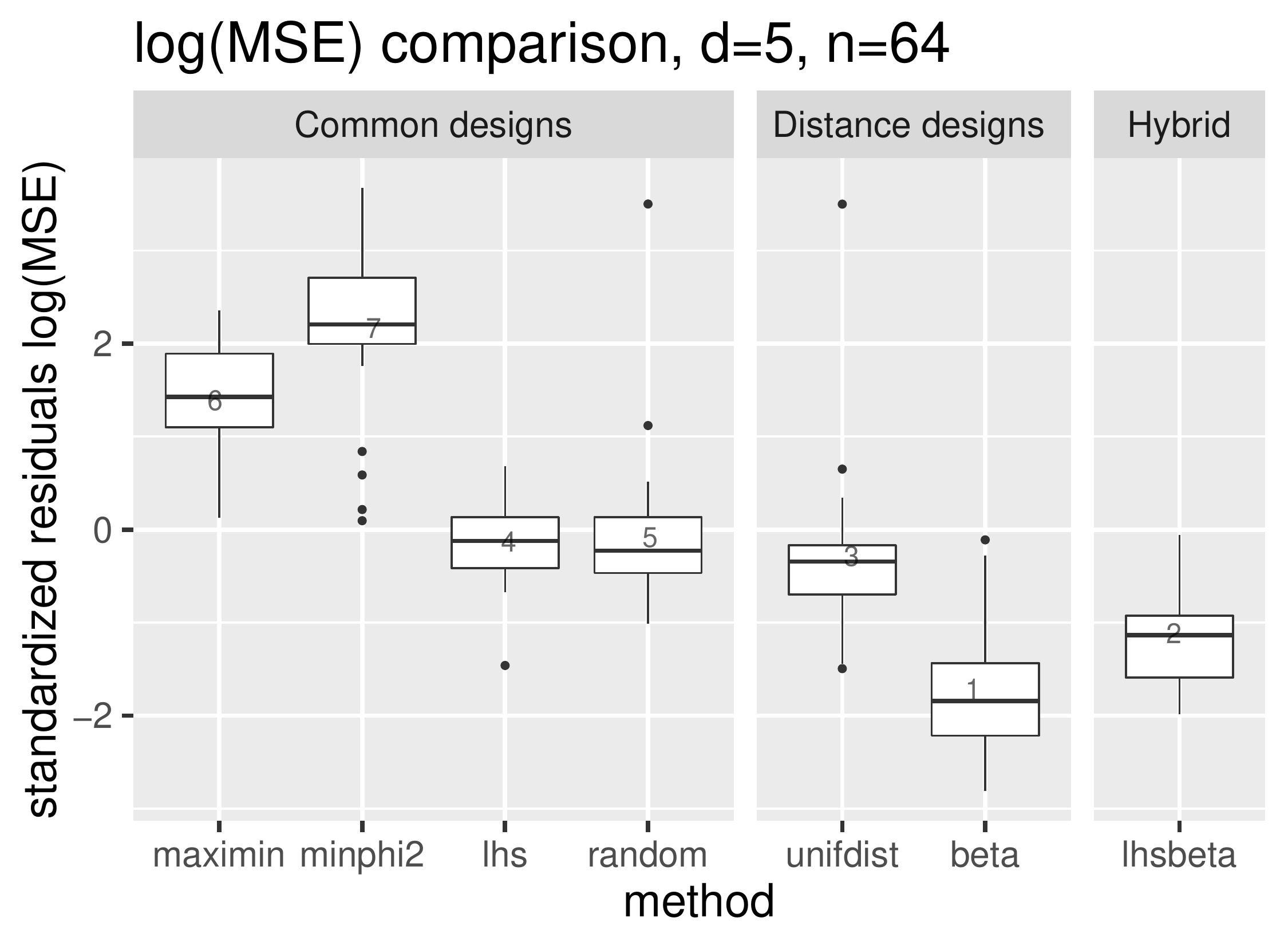}
\end{minipage}\quad

\begin{minipage}[b]{0.48\linewidth}
	\centering
	\includegraphics[width=\textwidth]{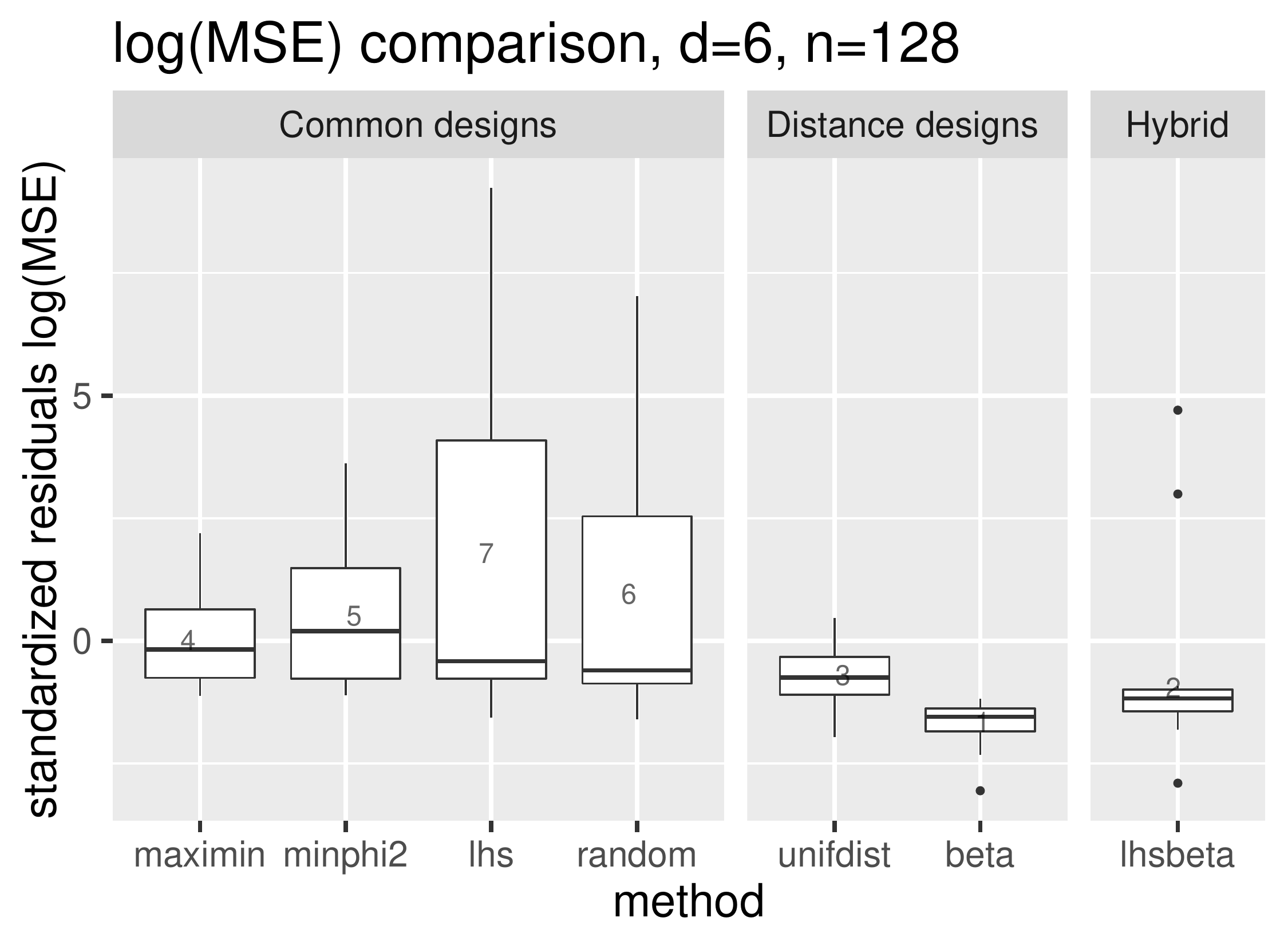}
\end{minipage}%
	\begin{minipage}[b]{0.48\linewidth}
		\centering \hspace{8pt}
		\resizebox{\columnwidth}{!}{
			\begin{tabular}{l|rrrrr}
        rank & 2d & 3d & 4d & 5d & 6d\\ \hline
			
				2 &  \textbf{9.78e-4}& 0.406            & \textbf{2.58e-4} & \textbf{2.90e-5} & \textbf{5.58e-3} \\ 
				3 & \textbf{1.29e-4} & \textbf{7.92e-8} & \textbf{1.16e-7} & \textbf{4.61e-7} & 0.213  \\ 
				4 & 0.388            & 0.151            & \textbf{7.14e-3} & 0.134            & \textbf{9.03e-8}  \\ 
				5 & \textbf{3.02e-3} & 0.287            & 0.415            & 0.446            & \textbf{6.64e-5} \\ 
				6 & \textbf{2.06e-5} & \textbf{1.08e-7} & \textbf{5.35e-4} & \textbf{3.51e-9} & 0.261 \\ 
				7 & \textbf{1.08e-5} & \textbf{0.0410}  & \textbf{4.72e-6} & \textbf{1.06e-6} & 0.123  \\ 
				
		\end{tabular}}
	\par\vspace{50pt}
	\end{minipage}
  \caption{Standardized logMSE boxplots to thirty gridded $\theta^{(t)}$
  values for seven comparators using $n=2^{d+1}$ over input dimension $d \in
  \{2,3,4,5,6\}$. The comparators are described in the text. \blu{Two
  outlying standardized $\log \mathrm{MSE}$ values were clipped by the
  $y$-axes to enhance boxplot viewing: random ($d=4$) at 10.9 and LHS
  ($d=6$) at 17.4.} The bottom-right panel provides $p$-values for lower-tail
  paired $t$-tests comparing adjacent performers as ranked by their mean
  logMSE from best (top) to worst (bottom).}
	\label{fig:theta}
\end{figure}

Figure \ref{fig:theta} shows boxplots of these standardized logMSEs,
marginalizing over $\theta^{(t)}$s, for all five experiments $d
\in \{2,3,4,5,6\}$.  The number written on each boxplot resides in the position of
the mean of that comparator, and indicates relative rank of that mean. In
order to help better quantify relative comparisons, the final panel provides
the outcome of pairwise paired $t$-tests, with pairing determined by adjacent
ranks: best vs.~second best, etc.  First consider the ``Common designs'' block
including boxplots of logMSEs for maximin, \blu{minphi2 ($\phi_2$)}, LHS and
random designs.  Although the final panel does not include a $p$-value for LHS
or random vs.~maximin \blu{when $d$ = 2,3}, because neither is ranked adjacently with maximin,
it is quite clear these beat maximin, \blu{which consistently beats minphi2}.
LHS and random, on the other hand, offer quite similar results.

\blu{Observe that the four ``Common designs'' follow a similar ranking for all
$d \leq 5$. However when $d=6$ maximin and minphi2 are better than LHS and
random. This happens because maximin's (and $\phi_p$'s) pathologies are partly
corrected in higher dimension. These designs push sites to the corners of the
input hyperrectangle.  As dimension grows the diversity of distances between
corners increases. This helps MSE, but only coincidentally.  Deliberate
diversity via unifdist and betadist is still better. }

The outcome of this experiment, including just those four common designs as
comparators, sparked our search for alternatives.  It is perhaps surprising
that a purely random design is at least as good for hyperparameter estimation
as more thoughtful alternatives like maximin and LHS.  The following
subsections describe our journey towards improved designs, ultimately
outlining details behind the other comparators in Figure \ref{fig:theta}.

\subsection{Uniform to beta designs}

\begin{figure}[ht!]
\centering
\includegraphics[width=3.5in,trim=0 10 0 40]{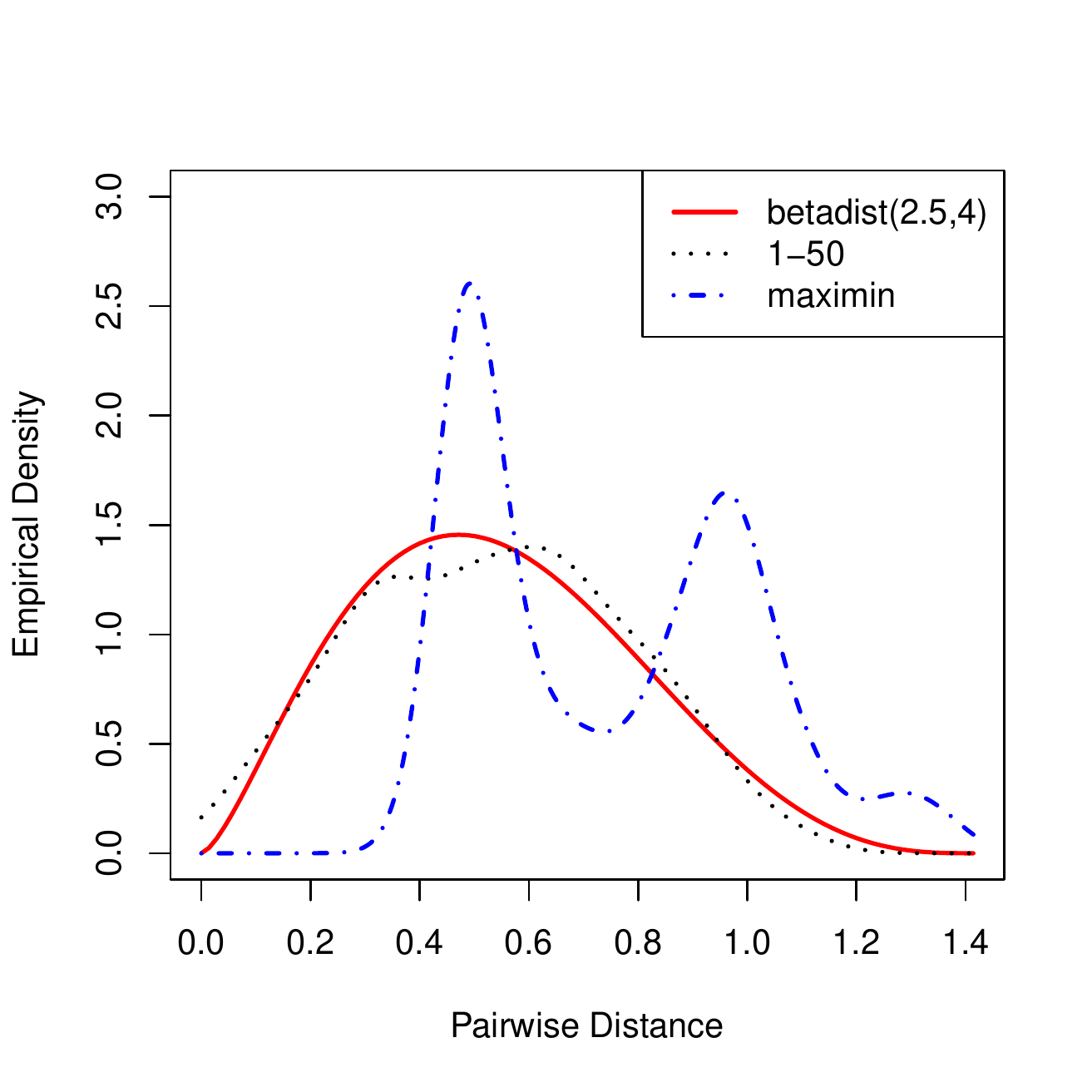}
\caption{Empirical density curves corresponding to random designs in 2d with
lowest 50 logMSE($\theta$) values from 1000 random design realizations.
Empirical maximin and $\mathrm{Beta}(2.5,4)$ densities are shown for
comparison.}
\label{fig:distdens}
\end{figure}

Intuitively, random and LHS are better than maximin for lengthscale ($\theta$)
inference because they result in a less adversarial distribution of pairwise
distances.  Maximin designs are calculated to ensure there are no small
pairwise distances, which is presumably too few. Consequently, the distance
distribution is multimodal: there are many distances near that minimum, with
the rest occurring at ``lower harmonics'' (multiples of that minimal
distance). Figure \ref{fig:distdens} offers a visualization.  Random and LHS
designs do not preclude small relative distances, although the latter does
enforce a degree of uniformity in position.  Both tend to yield distance
distributions which are unimodal. Figure \ref{fig:distdens} demonstrates this
for a subset of random designs, which will be discussed in more detail
momentarily.  The situation is similar for LHS, which we shall revisit in
Section \ref{sec:lhsbeta}.

\begin{algorithm}[ht!]
\textbf{Init:} Fill $\vecX$ with a random design of size $n$, i.e,
$\mathbf{x}_i\stackrel{\mathrm{iid}}{\sim}\mathrm{Unif}[0,1]^d$, $i=1,\dots,n$. \\
\For{$s = 1,\dots, S$}{
  Select an index $i\in \{1,\dots,n\}$ at random. \\
  Generate $\mathbf{x}_i' \sim  \mathrm{Unif[0,1]}^d$.\\
  Propose new design $\vecX'$ as $\vecX$ with $\mathbf{x}_i$ swapped with $\mathbf{x}_i'$.\\
  \If{$\mathrm{KSD}(\vecX', F) < \mathrm{KSD}(\vecX, F)$ }{
   $\mathbf{x}_i \leftarrow \mathbf{x}_i'$ in the $i^\mathrm{th}$ row of $\vecX$;\\
   i.e., accept $\vecX \leftarrow \vecX'$}
 }
\textbf{Return:} $n \times d$ design $\vecX$.\\
\vspace{0.5cm}
\caption{MC calculation of size $n$ in $[0,1]^d$ targeting distance distribution $F$.}
\label{alg:distdesign}
\end{algorithm}

The experimental outcomes just described  got us thinking about desirable
distance distributions for lengthscale estimation.  We speculated that it
could be advantageous to have a uniform distance design (unifdist), so that
all distances were represented---or as many as possible up to the desired
design size $n$.  Throughout we presume inputs have been scaled to $[0,1]^d$,
and restrict the search for lengthscales to $\theta \in (0, \sqrt{d}]$.  So
when we say uniform, or any other distribution, we mean
$\mathrm{Unif}(0,\sqrt{d}]$.\footnote{Our MLE calculations restrict $\theta$
to be greater than the square-root of machine precision, which is near {\tt
1e-8} on most machines.}  To calculate a design whose  distribution of
pairwise distances resembles a reference $F$, we follow the pseudo-code
provided by Algorithm \ref{alg:distdesign} which is based on $S$ stochastic
swap proposals that are accepted or rejected via Kolmogorov-Smirnov distances
(KSD) against $F$. In our examples we fix $S=10^5$ and utilize a
\blu{faster}, custom implementation of KSD based on \blu{isolating} the
\verb!$statistic! output of the built-in \verb!ks.test! function in {\sf R}.
\blu{Besides being stochastic, the search is greedy which means that it only
guarantees local convergence as $S \rightarrow \infty$.  Nevertheless we find
that in practice it furnishes empirical pairwise distance distributions close
to the target $F$.  There is little benefit in restarting the algorithm to search for a more
global optimum.}  

Unfortunately, our intuition about unifdist designs didn't completely match our results.  As
summarized along with our earlier RMSE comparison in Figure \ref{fig:theta},
unifdist designs are better than maximin, but worse than LHS and random. This
outcome prompted a more careful investigation into why random designs, work so well.

Consider the lines in Figure \ref{fig:distdens} labeled ``1--50'',
representing the empirical density of distances among the random designs whose
logMSE was among the fifty best in a large Monte Carlo (MC) exercise.  Observe
\blu{that this density is} unimodal, having more small distances than maximin and very
few really large distances.  The solid red curve in the figure is a
$\mathrm{Beta}(2.5,4)$ density scaled to $[0,\sqrt{2}]$ as a representative
example of a parametric distribution similar to that of those best random
distances.

Unifdist designs, which are not shown in the figure, target a flat line across
the $[0,\sqrt{2}]$ domain.  Unifdist outperforms maximin, but not the
best (or even the typical) random designs. This suggests that while having more
short distances is desirable, having as many distances at the extremes---both
large and small---may not be helpful on average.  As the results in Figure
\ref{fig:theta} show, having Beta-distributed distances, focusing the
distribution on mid--low-range pairwise distances, leads to statistically
significant improvements over random in all three cases.  In fact, these
``betadist'' designs (being ranked 2 or 1) are the only ones in that figure
whose logMSEs are statistically better (see $p$-values in the lower-right panel)
than all other designs of lower rank. 

Although Figure \ref{fig:distdens} suggests that a $\mathrm{Beta}(2.5,4)$ is a
good target distribution for a betadist design, that was not the specification
used \blu{to generate all results summarized} in Figure \ref{fig:theta}.
\blu{The best} setting of shape parameters, $(\hat{\alpha}, \hat{\beta})$ in
$\mathrm{Beta}(\alpha, \beta)$ depends on dimension $d$ and design size $n$,
as we explore below. However, it is worth nothing that $\mathrm{Beta}(2.5,4)$
does generally perform well \blu{because, as we show, the set of decent
$(\alpha, \beta)$ values is relatively big, and does not vary substantially in
$n$ and $d$.  But it's not so big as to choose arbitrarily.}

\subsection{Optimization of shape parameters of betadist design}
\label{sec:betaopt}

Here we view the choice of betadist parameterization, $\hat{\alpha}$ and
$\hat{\beta}$ in $\mathrm{Beta}(\alpha, \beta)$, for particular design size
$n$ in input dimension $d$, as an optimization problem. \blu{I.e., we wish to
automate the search for $\mathrm{betadist}_{n,d}(\hat{\alpha}, \hat{\beta})$.
Discussion around Figure \ref{fig:rmses} indicates that a degree of detrending
will be required in order to not over-emphasize larger $\theta$ settings in
the optimization criteria.  To address this, we seek $(\hat{\alpha},
\hat{\beta}) = \mathrm{argmin}_{\alpha, \beta} \;
\mathrm{deRIMSE}_{n,d}(\alpha, \beta)$ where the criteria deRIMSE is defined
following a scheme similar to that described around Figure \ref{fig:rmses}.}

\blu{Begin by establishing a regular grid of $\theta$ values $(\theta^{(1)}=0.1,
\dots, \theta^{(T)} =
\sqrt{d})$, just like in Figure \ref{fig:rmses}.  Next, generate one pair 
$(\alpha, \beta) \sim \mathrm{Unif}(1,10)^2$ and use these to create $D$
designs $\mathbf{X}_n^{(i)} \sim \mathrm{betadist}_{n,d}(\alpha, \beta)$, for
$i=1,\dots,D$ following Algorithm \ref{alg:distdesign}.  Averaging over more
random $(\alpha, \beta)$ will be described momentarily. For each
$\mathbf{X}_n^{(i)}$ and each $\theta^{(t)}$ generate random responses
$\mathbf{Y}_n^{(t,i)} \sim (\mathbf{X}_n^{(i)}, \theta^{(t)})$ under the GP MVN
and estimate $\hat{\theta}^{(t,i)}$ via MLE.  Finally, calculate
$\mathrm{RMSE}^{(t)} = \sqrt{\sum_{i=1}^D (\hat{\theta}^{(t,i)} -
\theta^{(t)})^2/D}$ to estimate the accuracy of those MLE calculations for each
$t=1,\dots,T$.  Then draw new $(\alpha,\beta) \sim
\mathrm{Unif}(1,10)^2$ yielding $\mathrm{RMSE}^{(t,r)}$, repeating the entire
scheme above $R$ times, i.e., for $r=1,\dots,R$. In our empirical work, we
chose $D=5$ and $R = T = 30$.}

\blu{Next, take pairs $(\theta^{(t)}, \{\mathrm{RMSE}^{(t,r)}\}_{r=1}^R)$ as $T
\times R$ observations of the quality of lengthscale estimation -- RMSE
dynamics -- across $\theta$-space and fit a Student-$t$ {\tt hetGP} to these
observations yielding a surrogate described by mean $\mu_t \equiv
\mu(\theta^{(t)})$ and $\sigma^2_t \equiv \sigma^2(\theta^{(t)})$.  Now we
are ready to define the criteria $\mathrm{deRIMSE}_{n,d}(\alpha, \beta)$ as
\[
\mathrm{deRIMSE}(\alpha, \beta) \equiv 
\frac{1}{T}\sum_{t=1}^{T} \frac{\mathrm{RMSE}^{(t)}(\alpha, \beta) - \mu_t}{\sigma_t}, 
\]
where $\mathrm{RMSE}^{(t)}(\alpha, \beta)$ is calculated just as described
above with the specific (not random) settings of $(\alpha, \beta)$ in question.}

As a warmup experiment toward solving that optimization problem, consider
$n=16$ and $d=2$.  We built a size 200 LHS design of $(\alpha, \beta)$
settings in $[1,10]^2$ with 5 replicates on each for a total of 1000
evaluations of $\mathrm{deRIMSE}$.  The bottom end of that
region, $\alpha,\beta \geq 1$, was chosen to limit the search to unimodal beta
distributions; the top end of 10 was chosen based on a smaller pilot study.
Each deRIMSE evaluation took about 50 seconds, leading to almost 14
hours of total simulation time.

\begin{figure}[ht!]
\centering
\includegraphics[width=3.5in,trim=0 0 0 40]{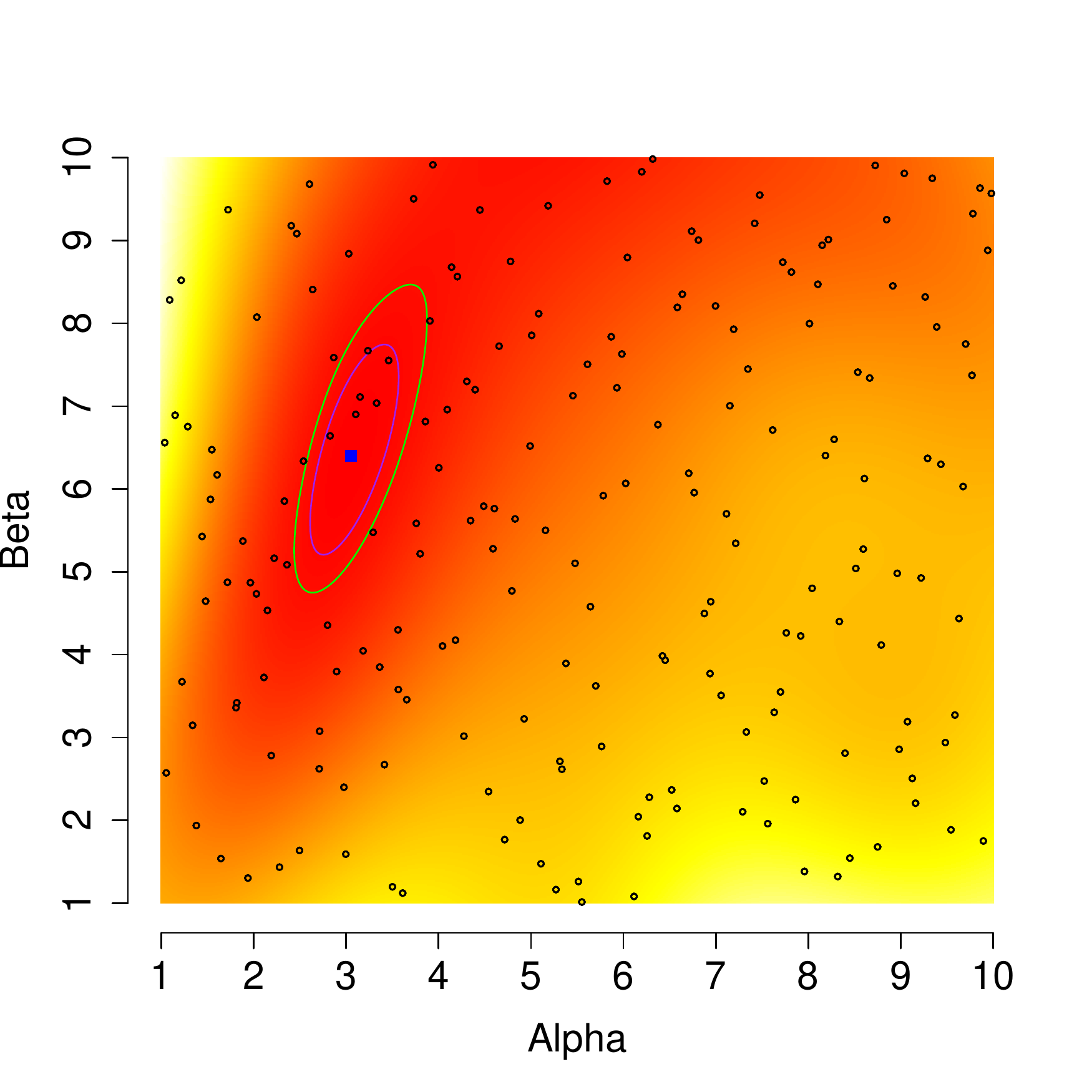}
\caption{deRIMSE surface with $T=1000$ for $n=16$ and $d=2$ as estimated by
{\tt hetGP}.  Dots show the design sites; lighter (heat) colors
correspond to higher deRIMSEs.}
\label{fig:GP216}
\end{figure}

Figure \ref{fig:GP216} shows the design (dots) and fitted surface of deRIMSE
values obtained with {\tt hetGP}, i.e., treating $\mathrm{deRIMSE}$ simulation
as a stochastic computer experiment and fitting a surrogate to a limited
number of evaluations.  Outliers are less of a concern when averaging over
$\theta^{(t)}$-values, so there was no need to include Student-$t$ features in
this regression. However, accommodating a degree of heteroskedasticity and
leveraging replication in the calculations were essential to obtain a good fit
in a reasonable amount of time \citep{hetGP1}.  The blue square, at about
$(\hat{\alpha}, \hat{\beta})=(3, 6.5)$ in the figure, shows where the
predictive surface is minimized; the green and purple contours outline regions
wherein predicted deRIMSE values are within 5\% and 10\% of that best setting.

Fourteen hours of simulation in order to choose the characteristics of a
random design is rather extreme.  However, once done for a particular choice
of covariance structure, design size $n$ and dimension $d$, it need not be
re-done.  Still, finding appropriate designs in higher dimension, with more
runs to fill out the larger volume could be computationally daunting. Doubling
$n$, for example, would result in more than double the computational effort.

For a more thrifty approach we turn to BO via EI.  The idea is to replace
a space-filling evaluation with a sequential design strategy that targets the
minimum of the mean of $\mathrm{deRIMSE}$. For a given $(n,d)$-setting, the
setup is as follows.  Begin by performing deRIMSE calculations on a maximin
design of size twenty, with ten replicates at each setting, and by fitting a
{\tt hetGP} to those realizations, deriving a predictive surface.  Then comes
the so-called BO acquisition.  Based on {\tt hetGP} posterior predictive
equations described by mean $\mu(\mathbf{x})$ and standard deviation
$\sigma(\mathbf{x})$, where $\mathbf{x}=(\alpha, \beta)$ in this case,
numerically optimize $\mathrm{EI}(\mathbf{x})$:
\begin{equation}
\mathrm{EI}(\mathbf{x}) = (\mu_{\min} - \mu(\mathbf{x})) 
\Phi\left(\frac{\mu_{\min} - \mu(\mathbf{x})}{\sigma(\mathbf{x})} \right) + \sigma(\mathbf{x}) 
\phi \left(\frac{\mu_{\min} - \mu(\mathbf{x})}{\sigma(\mathbf{x})}\right), \label{eq:ei}
\end{equation}
where $\mu_{\min} = \min_\mathbf{x} \mu(\mathbf{x})$ and $\Phi$ and $\phi$ are
the standard Gaussian cdf and pdf, respectively. After (a) solving
$\mathbf{x}^* = \mathrm{argmin}_\mathbf{x} \; \mathrm{EI}(\mathbf{x})$, which
we accomplish using a hybrid of discrete search over replicates and continuous
multi-start {\sf R}--{\tt optim}-based search with
\verb|method="L-BFGS-B"|; (b) simulating $y^* = \mathrm{deRIMSE}(\mathbf{x}^*)$;
and (c) incorporating the new data pair into the design and updating the {\tt
hetGP} model fit; the process repeats (back to (a)).

For details on EI and BO see \cite{jones:schonlau:welch:1998} and Chapter 6.3
of \cite{santner:etal:2003}. \citet{snoek:etal:2012} offer a somewhat more
modern machine learning perspective centered around the use of BO for
estimating hyperparameters of deep neural networks.  Our use here---to tune a
design---is related in spirit but distinct in form. In fact, the setup we
propose is fractal.  It solves a design problem (for estimating lengthscale)
with the solution to another design problem: for function minimization.   One
could argue that our choice of an initial maximin design for BO is
sub-optimal, and we will do just that in Section \ref{sec:ei}.


\begin{figure}[ht!]
\centering
\small
\tabcolsep=0.11cm
\begin{tabular}{ccc}

\includegraphics[trim=30 20 0 60,width=51mm]{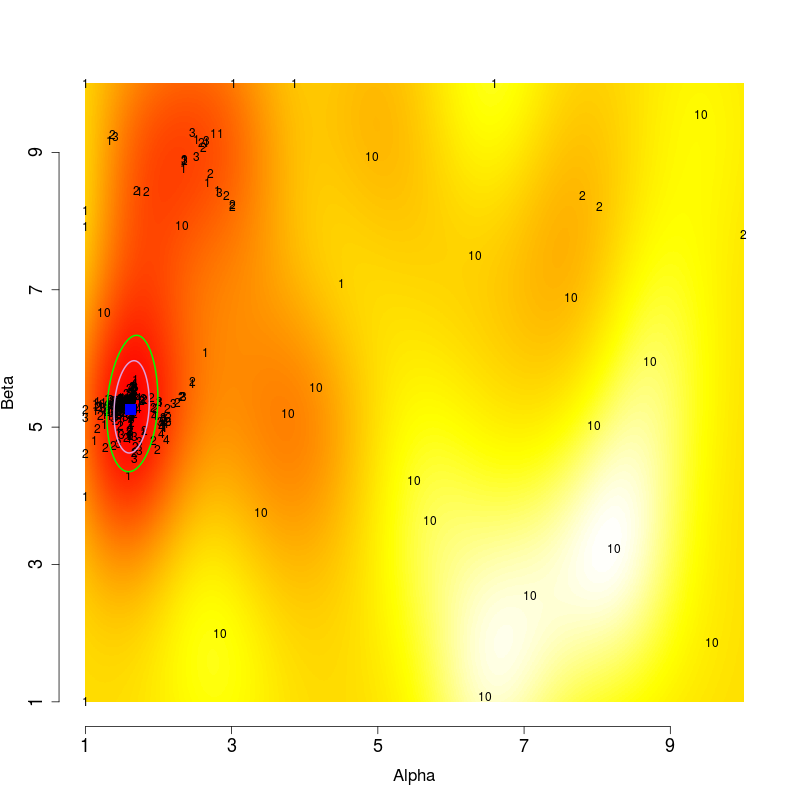} & 
\includegraphics[trim=30 20 0 50,width=51mm]{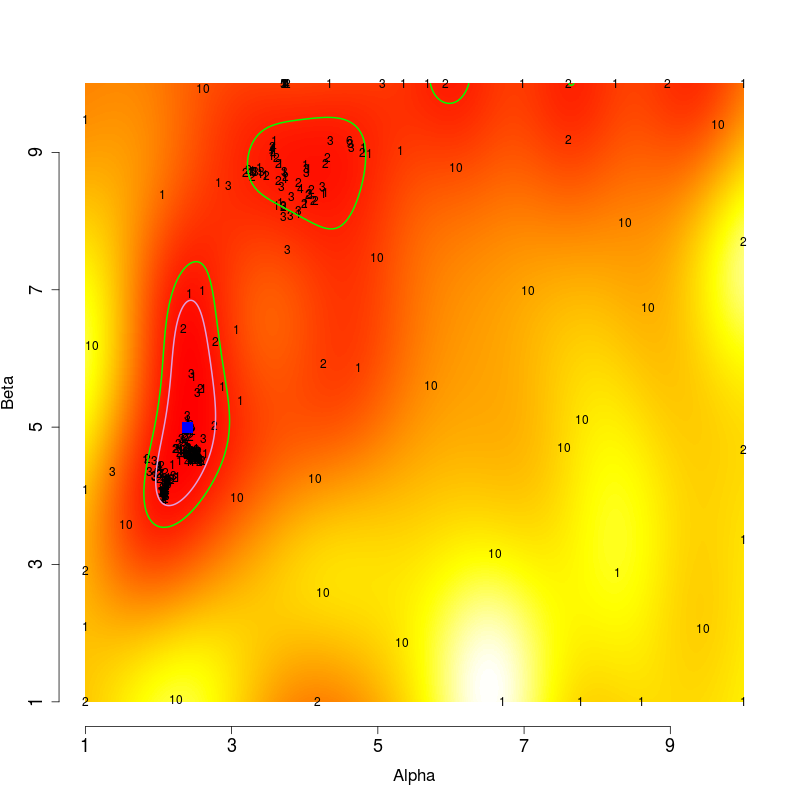} &  
\includegraphics[trim=30 20 0 50,width=51mm]{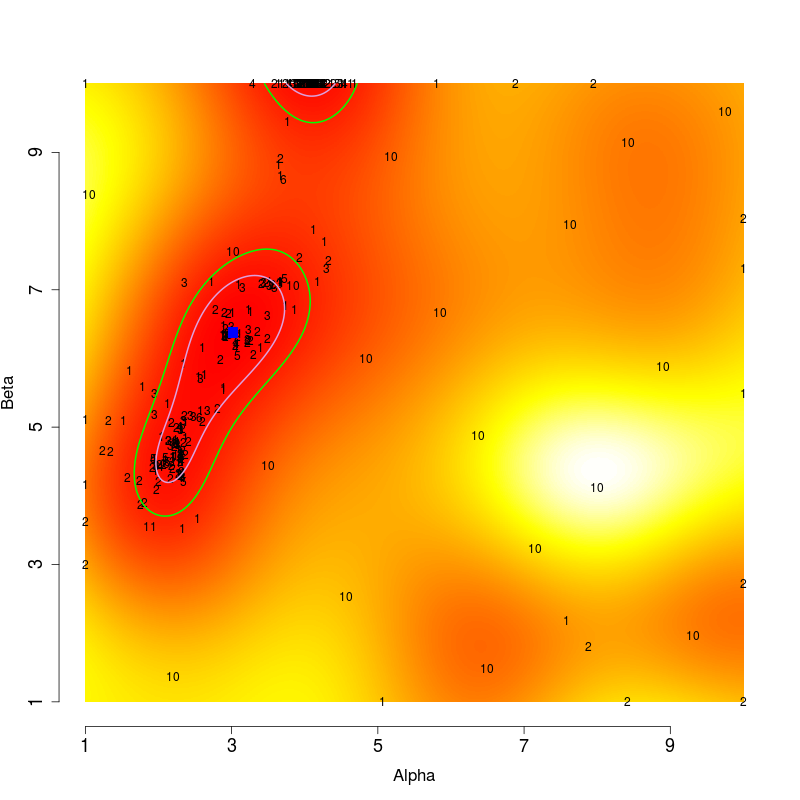} \\
$n=8$, $d=2$ & $n=16$, $d=2$ & $n=16$, $d=3$ \\
$\hat{\alpha}=1.62$, $\hat{\beta}=5.26$ & $\hat{\alpha}=2.4$, $\hat{\beta}=5$ & $\hat{\alpha}=3.02$, $\hat{\beta}=6.38$ \\ 
$\tilde{\alpha}=1.5$, $\tilde{\beta}=5$ & $\tilde{\alpha}=2$, $\tilde{\beta}=4$ & $\hat{\alpha}=2.5$, $\tilde{\beta}=5$ \\
\includegraphics[trim={30 20 0 15},width=51mm]{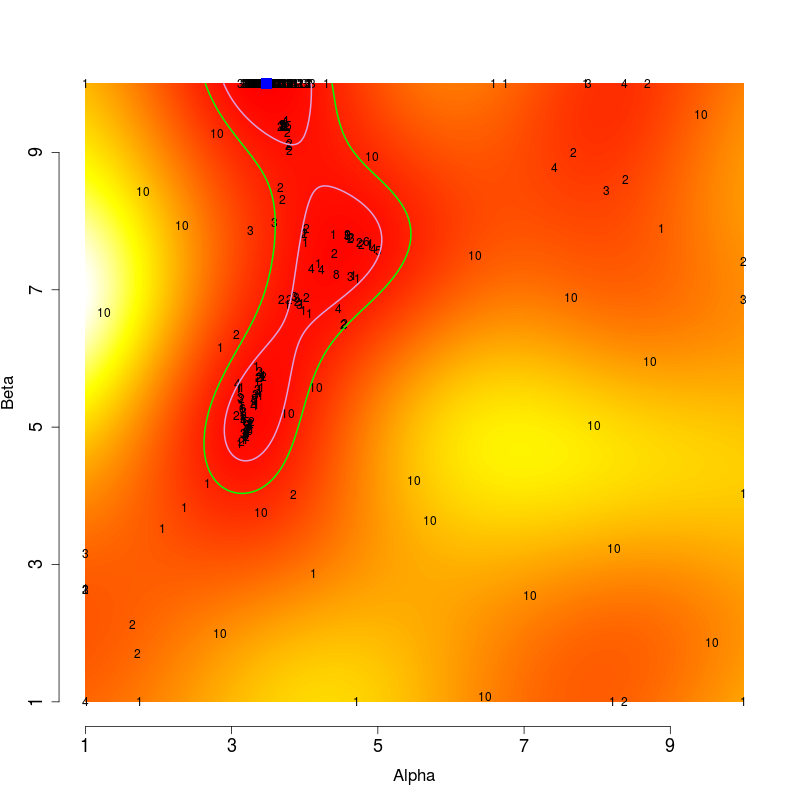} & 
\includegraphics[trim={30 20 0 15},width=51mm]{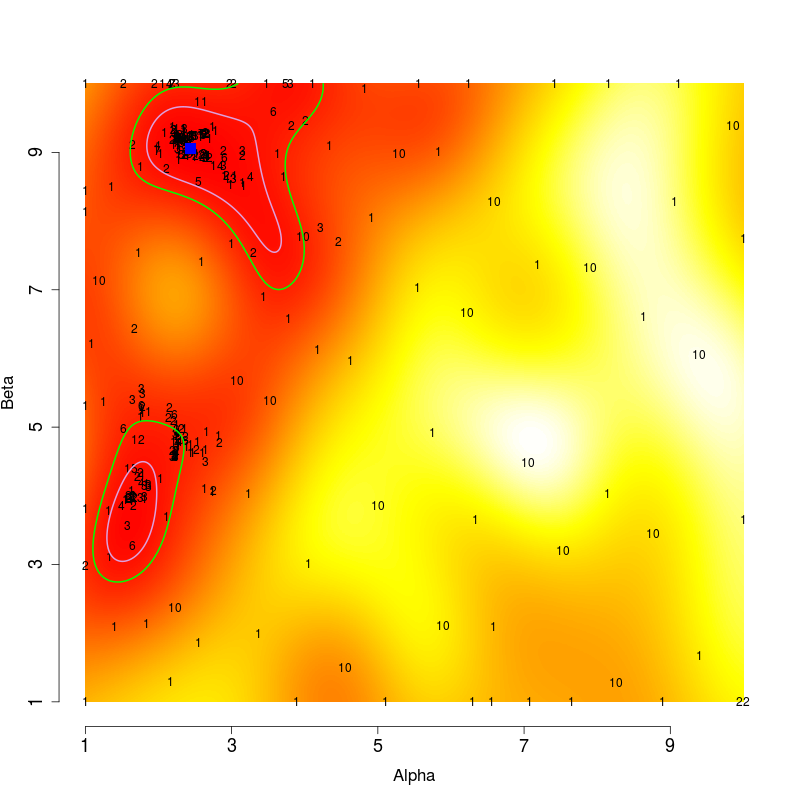} &  
\includegraphics[trim={30 20 0 15},width=51mm]{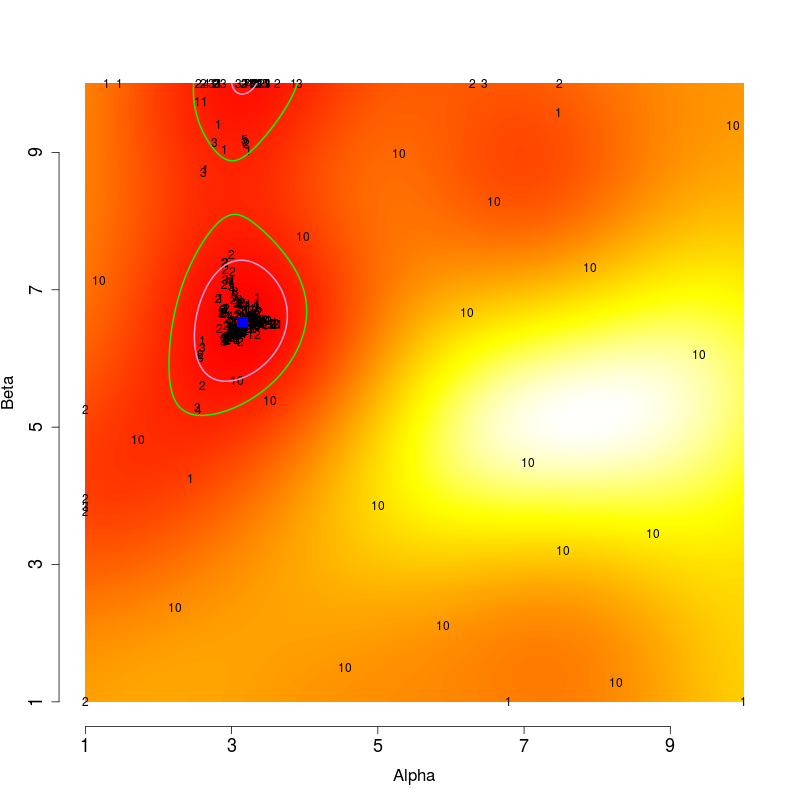} \\
$n=32$, $d=3$ & $n=32$, $d=4$ & $n=64$, $d=4$ \\
$\hat{\alpha}=3.48$, $\hat{\beta}=10$ & $\hat{\alpha}=2.44$, $\hat{\beta}=9.06$ & $\hat{\alpha}=3.15$, $\hat{\beta}=6.53$ \\
$\tilde{\alpha}=3$, $\tilde{\beta}=5$ & $\tilde{\alpha}=1.5$, $\tilde{\beta}=3.5$ & $\hat{\alpha}=3$, $\tilde{\beta}=6$ \\
\includegraphics[trim={30 20 0 15},width=51mm]{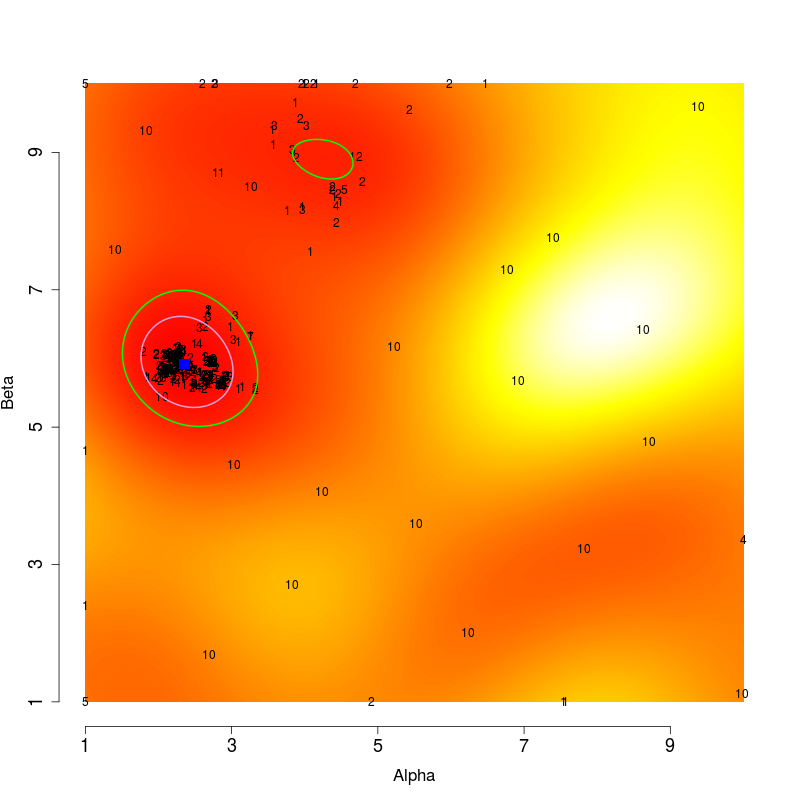} & 
\includegraphics[trim={30 20 0 15},width=51mm]{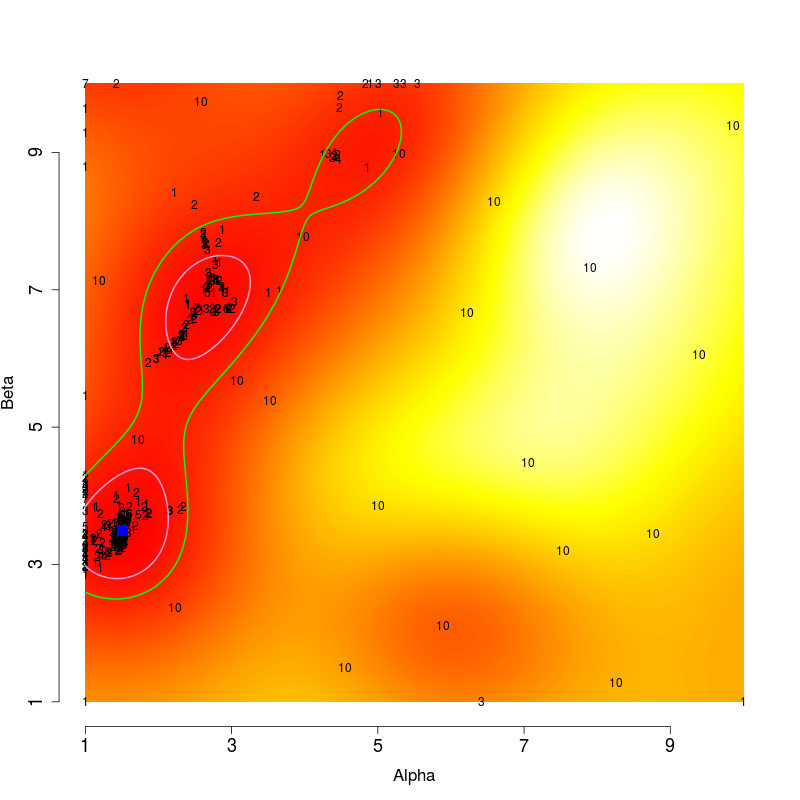} &  
\includegraphics[trim={30 20 0 15},width=51mm]{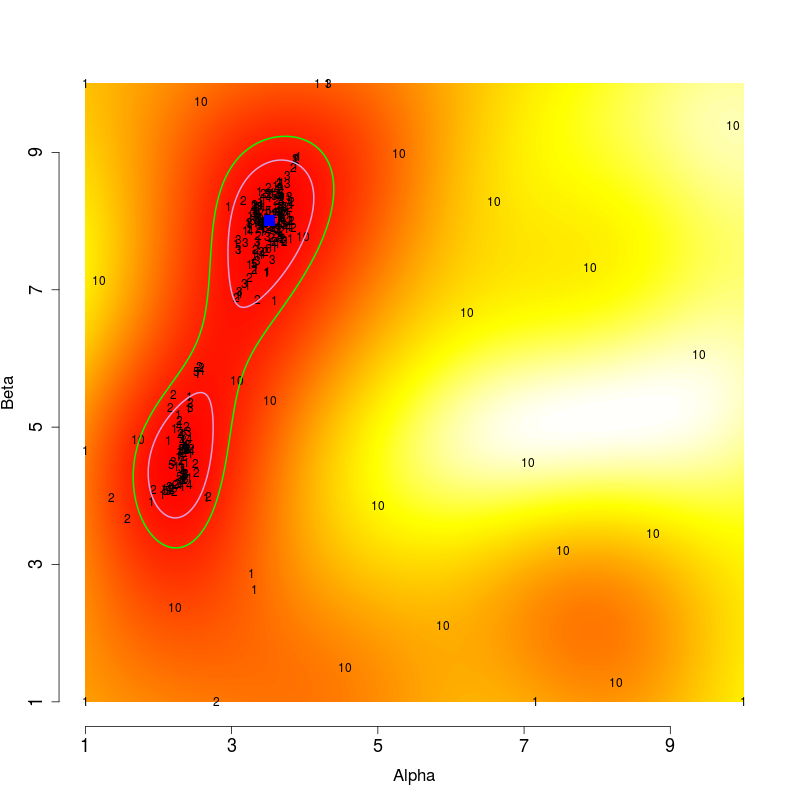} \\
$n=64$, $d=5$ & $n=128$, $d=5$ & $n=128$, $d=6$ \\
$\hat{\alpha}=2.36$, $\hat{\beta}=5.92$ & $\hat{\alpha}=1.51$, $\hat{\beta}=3.5$ & $\hat{\alpha}=3.52$, $\hat{\beta}=8.01$ \\
$\tilde{\alpha}=2$, $\tilde{\beta}=6$ & $\tilde{\alpha}=1$, $\tilde{\beta}=3$ & $\hat{\alpha}=2$, $\tilde{\beta}=4$ \\

\end{tabular}
\caption{Outcomes of BO of RIMSE surfaces for various choices of $n$ and $d$.
Numbers show location and number of replicates in acquisitions; blue square
shows $(\hat{\alpha}, \hat{\beta})$; purple and green contours show 5\% and 10\% from the optimal.}
\label{f:betaopt}
\end{figure}

For a set of representative $n$ and $d$, we allowed our BO scheme to collect
an additional $600$ deRIMSE simulations. The resulting selections, overlayed
with final predictive mean surface from {\tt hetGP}, the best value of
$(\hat{\alpha}, \hat{\beta})$ and a 5\% and 10\% contour are shown in Figure
\ref{f:betaopt}.  Several noteworthy patterns emerge from the panels in the
figure.  First, although some of the surfaces appear to be multimodal, or at
least to have ridges of low deRIMSE values, there is usually a setting with
relatively low $(\alpha, \beta)$  which works well.  Sometimes a larger
setting is predicted as optimal; but there is usually an alternate setting,
reported as $(\tilde{\alpha}, \tilde{\beta})$ in the figure, which is almost
as good (within 5\%).

These ``near-optimal'' $(\tilde{\alpha}, \tilde{\beta})$ were used in our
betadist designs, and subsequent boxplots and $p$-value calculations, in
Figure \ref{fig:rmses}.  They are re-used throughout the remainder of the
paper in our empirical work [Sections \ref{sec:alm}--\ref{sec:ei}], and
likewise with the hybrid lhsbeta designs discussed momentarily.  Although the
computational demands are still sizable even with the more thrifty BO, these
designs are ``up-front''.  Once saved as we do for the nine choices above,
no recalculation is required.

\section{Hybrid betadist and LHS}
\label{sec:lhsbeta}

Having a betadist design, which provides better estimates of hyperparameters
like the lengthscale $\theta$, is advantageous only insofar as the resulting
surrogate fits, i.e., their predictive equations (\ref{eq:predgp}), are
accurate. Since GP surrogates are inherently spatial predictors, practitioners
have long preferred designs which fill the space, so that those sites may
serve as nearby anchors to good out-of-sample predictive performance. Betadist
designs space-fill less than common alternatives, both quantitatively (i.e.,
via the maximin criteria) and qualitatively (since they're inherently random).
Thus they hold the potential to be inferior as predictive anchors. Yet in our
empirical work, we've only been able to demonstrate this negative result (not
shown here) when good hyperparameter settings are known.  Betadist shines
brightest in sequential application [Section \ref{sec:seq}], where the impact
of early estimates of hyperparameters can have a substantial
affect---exceptionally deleterious in pathological cases---on subsequent
design decisions in several common situations.

Still, betadist designs consider only relative distance, completely ignoring
position except that the points lie in the study area.  Among more-or-less
equivalent optimal betadist designs, some may have better positional
properties and thus offer better anchoring for prediction without compromising
on hyperparameter quality.  To explore this possibility we considered a hybrid
between betadist and LHS designs.  Our ``lhsbeta'' is similar in spirit to
maximin--LHS hybrids where maximin helps avoid second-order aliasing common
with LHSs, and LHS helps maximin avoid clumpy marginals. In
lhsbeta, we primarily view LHS as helping betadist acquire a degree of
positional preference, however the alternate perspective of preferring LHSs
with better relative distances is no less valid.

Our stochastic search strategy for finding lhsbeta designs is coded
in Algorithm \ref{alg:lhsbeta}.
\begin{algorithm}[ht!]
\textbf{Init:} Fill $\mathbf{X}$ with a LHS of size $n$ in $d$ dimensions.\\
\For{$s=1,\dots,S$}{
  Randomly select a pair of design points $\mathbf{x}_i, \mathbf{x}_j$. \\
  Randomly select a dimension $k \in \{1,\dots, d\}$. \\ 
  Propose a new design $\mathbf{X}'$ by swapping Latin squares $L_{i,k}$ and $L_{j,k}$ producing \\
  \ \ new $\mathbf{x}_i'$ and $\mathbf{x}_j'$ after re-jittering with $2d$ new uniform random numbers.\\
  \If{$\mathrm{KSD}(\mathbf{X}', F) < \mathrm{KSD}(\mathbf{X}, F)$ }{
  	$\mathbf{x}_i \leftarrow \mathbf{x}_i'$ and $\mathbf{x}_j \leftarrow \mathbf{x}_j'$ in the $(i,j)^\mathrm{th}$ rows of $\mathbf{X}$;\\
  	i.e., accept $\mathbf{X} \leftarrow \mathbf{X}'$
  }
 }
\caption{Hybrid $F$-dist--LHS via $S$ MC iterations for a design of size $n$ in $d$ dimensions.}
\label{alg:lhsbeta}
\end{algorithm}
Like in Algorithm \ref{alg:distdesign} for betadist, we presume an input space
coded to $[0,1]^d$. The algorithm is initialized with an LHS $\mathbf{X}$,
built in the canonical way \citep[see, e.g.,][]{lin:tiang:2015} by first
choosing $d$ random permutations of $\{1,\dots,n\}$, saved in a $n
\times d$ matrix $\mathbf{L}$ describing the $n$ selected hypercubes out of
the $n^d$ possible partitions of the input space, and then applying jitter in
that selected cube.  Each subsequent iteration of stochastic search involves
randomly proposing to swap pairs of rows and columns of $\mathbf{L}$,
effectively swapping the pair of Latin squares without destroying the
one-dimensional uniformity property, and then rejittering that pair points
within their respective squares. That proposal is then accepted or rejected
according to KSD measured against a distribution $F$, which in our
applications is $\mathrm{Beta}(\tilde{\alpha}, \tilde{\beta})$ from Section
\ref{sec:randomtobeta}.  Since two types of random proposals are being
performed simultaneously, compared to Algorithm \ref{alg:distdesign}'s single
random swap, we prefer a multiple of two larger $S$ in Algorithm
\ref{alg:lhsbeta}; $S = 10^5$ in our empirical work.

\begin{figure}[ht!]
\centering
\includegraphics[scale=0.55,trim=32 40 20 0]{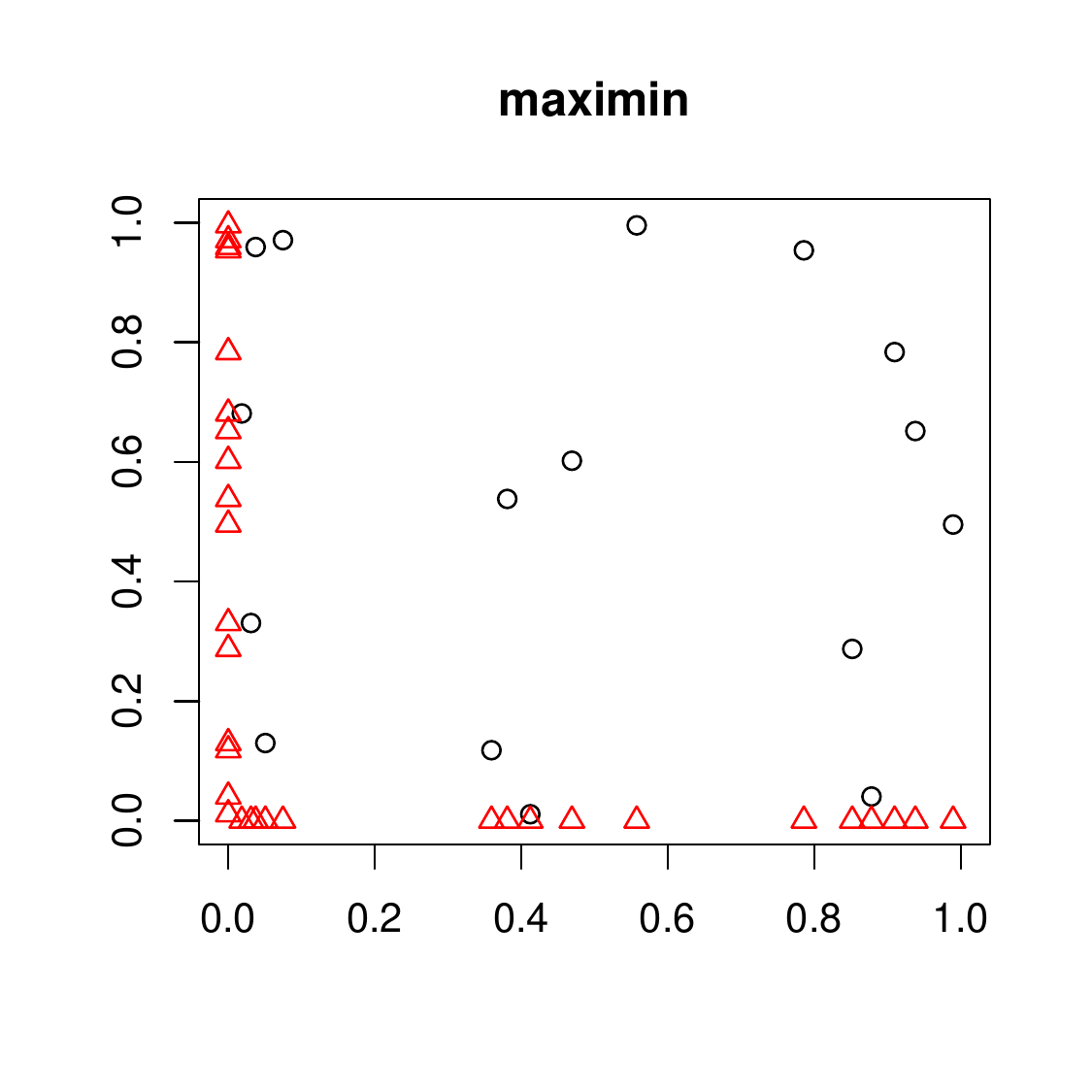}
\includegraphics[scale=0.55,trim=32 40 20 0]{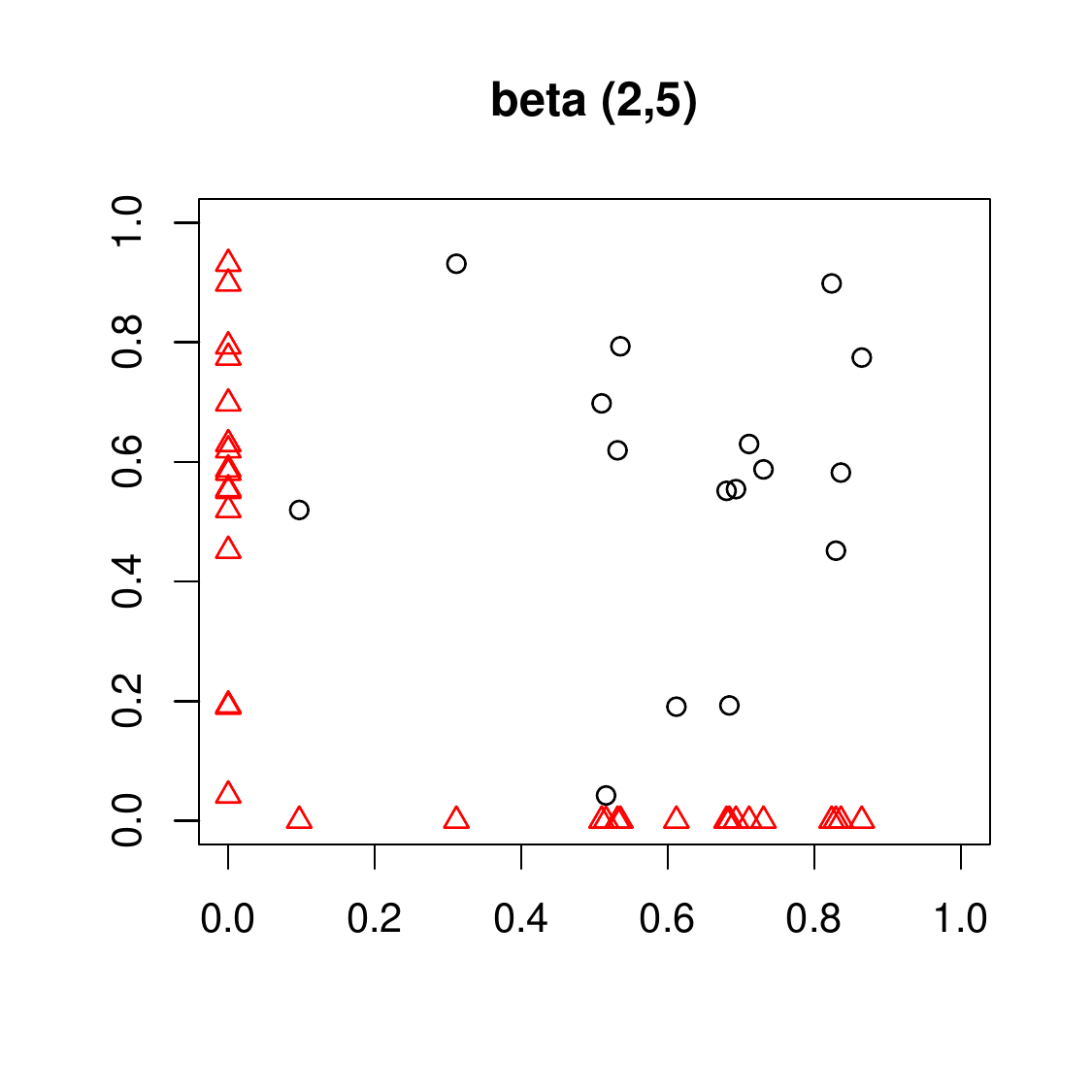}
\includegraphics[scale=0.55,trim=32 40 20 0]{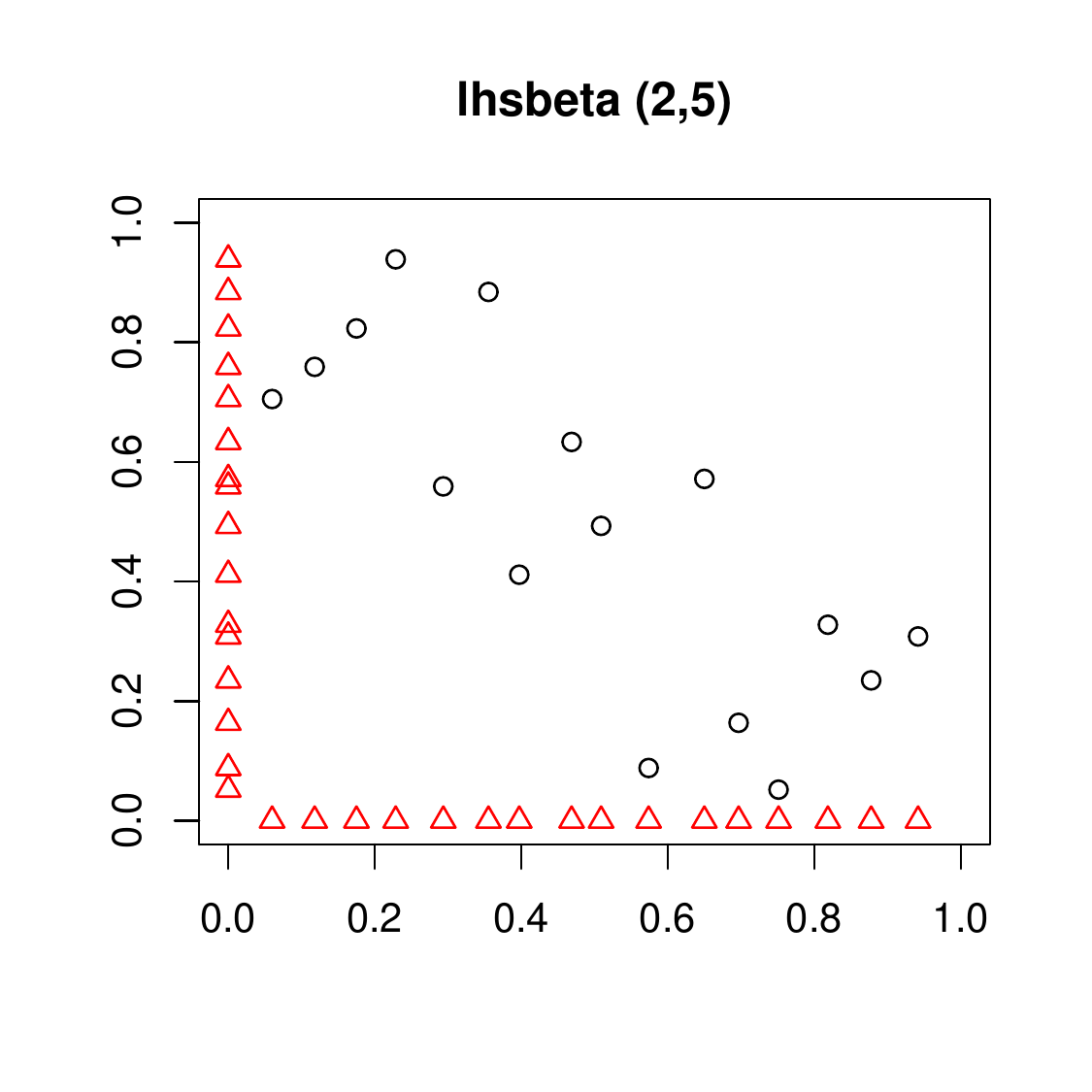}
\caption{2d (black circles) and 1d  (red triangles) projections of
three $d=3$ designs, $n=16$.}
\label{fig:designs}
\end{figure}

Figure \ref{fig:designs} shows a visual comparison between maximin, betadist
and lhsbeta designs so constructed.  The plots provide a 2d projection for the
case $n=16$ and $d=3$.  Observe that maximin's 1d margins, shown as red
triangles at the axes in the left panel, are not uniform.  Neither are those
in the 2d projection shown as open circles.  First-order aliasing is severe in
both projections.  In the middle panel, our betadist design has a similar
problem (although perhaps not to the same degree), yet we know that the
distribution of pairwise distances in 3d are much better than maximin for the
purpose of lengthscale inference.  In the right panel the 1d and 2d margins look
much better, because the sample is an LHS.  Among LHSs, this lhsbeta design
has a near optimal distribution of pairwise distances for this setting
$(n,d)$. \blu{Figure \ref{fig:theta} shows that lhsbeta designs are sometimes
worse than ordinary betadist
designs, but they both are consistently better than all of the other
comparators in the figure.}  This is perhaps not surprising because lhsbeta
designs are indeed betadist designs, yet selected for an additional feature
not relevant for lengthscale information: space-fillingness.  As we show in
two prediction-based comparisons below, lhsbeta designs are sometimes superior
on those tasks.

\section{Application to sequential design}
\label{sec:seq}

Here we provide two applications of betadist and lhsdist as initial designs
for a subsequent sequential analysis.  In both cases, these distance
distribution-based designs are only engaged in a limited way, as a means of
seeding the sequential procedure.  Subsequent design acquisitions are
then off-loaded to other criteria. Still, it is remarkable how profound the
effect of this initial choice can be.  A poorly chosen initial design of just
$n_{\mathrm{init}} = 8$ points, say, can be detrimental to predictive accuracy
at $n=64$.

\subsection{Active Learning MacKay}
\label{sec:alm}

First consider the so-called {\em active learning MacKay}
\citep[ALM;][]{mackay:1992} method of sequential design for reducing
predictive uncertainty.  Acquisitions are determined by maximizing the
predictive variance $\sigma^2(\mathbf{x})$.  The rationale for that choice is
that ME designs for GPs involve maximizing the determinant of the covariance
matrix $\mathbf{K}_n$, and one can show that $\log |\mathbf{K}_{n+1}| = \log
|\mathbf{K}_n| + \log \sigma^2(\mathbf{x}_{n+1})$. After many applications the
result is space-filling, however the degree to which design points are pushed
toward the boundaries of the study region depends crucially on the
lengthscale(s) used to define $\mathbf{K}_n$.  For more details on ALM with
GPs see, e.g., \citet{seo00} and \citet{gra:lee:2009}.

The target of our experiment is a function $f(\mathbf{x})$ observed under
light noise as $Y(\mathbf{x}) = f(\mathbf{x}) + \varepsilon$, with $\varepsilon
\stackrel{\mathrm{iid}}{\sim} \mathcal{N}(0, 0.01^2)$. For $f(\mathbf{x})$ we use
the function
\[
f(\mathbf{x}) = x_1 \exp\{-x_i^2 - x_2^2\} \quad \mbox{ with } 
\mathbf{x} \in [-2,4]^2,
\]
first introduced as an active learning benchmark by \citet{gra:lee:2009}. We
begin with an initial design of size $n_{\mathrm{init}} = 8$, and perform 56
additional ALM acquisitions for a total of $n=64$ evaluations.  Along the way,
root mean-squared prediction error (RMSPE) is calculated on noise-free outputs
obtained on a regular $100\times 100$ testing grid in the input space.  For the
initial design, we consider random, LHS, 2d optimal $(\tilde{\alpha}=2,
\tilde{\beta}=5)$ betadist and lhsdist designs, and maximin.  Unifdist has
been dropped from the comparison on the grounds that it is a sub-optimal
betadist alternative.

In keeping with our earlier experiments, MLE calculations limited to
$(0,\sqrt{2}]$ are updated after each sequential design acquisition.  To
accommodate the noisy evaluations, we augment our covariance with a nugget
hyperparameter which is included in the MLE calculation via {\tt jmleGP} in
the {\tt laGP} package.  An L-BFGS-B scheme is used to solve
$\mathrm{argmin}_\mathbf{x}
\sigma^2(\mathbf{x})$ via {\tt optim} in {\sf R}.  Variance surfaces can be
highly multi-modal, having as many maxima as design points which is what
creates the ``sausage''-like shape characteristic of the error-bars produced
by GP predictive equations.   We deployed an $n$-factor sequential maximin
multi-start scheme to avoid inferior local modes of the variance surface.
This means that maximin is used to choose the {\tt optim} initializations, in
order to space out starting locations relative to each other and to the
existing $\mb{X}_n$ design locations.

\begin{figure}[!ht]
\centering
\includegraphics[scale=0.6,trim=0 0 0 30]{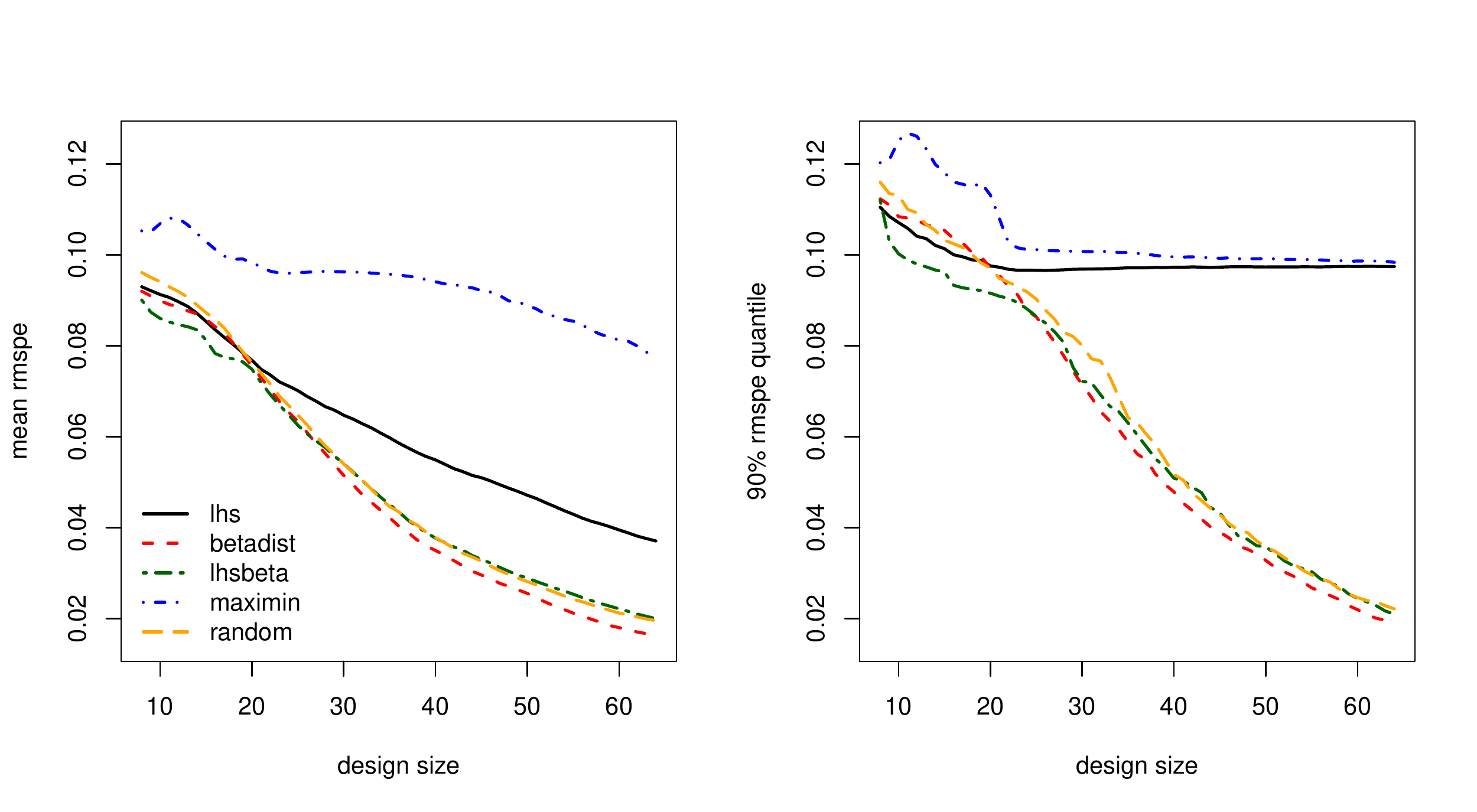}
\caption{RMSPE comparison of initial designs ($n_{\mathrm{init}} =8$) as a
function of the number of subsequent sequential design iterations via ALM.
Each comparator has a pair of lines: those in the left panel indicate mean
RMSE; those on the right are the upper 90\% quantile.}
\label{f:alm}
\end{figure}

Figure \ref{f:alm} shows the outcome of this exercise via mean RMSPE (left
panel) and upper 90\% RMSPE quantile (right) obtained from 1000 MC repetitions
of the scheme described above.  Several striking observations stand out.
Betadist, lhsbeta and random perform about the same, with betadist winning out
in the end.  However in early stages lhsdist is best and random is the worst
of the three. Beta-distributed distances (from betadist and lhsbeta) lead to
better hyperparameter estimates than random. Yet position of design sites is
more important than lengthscale quality when there are little data. After many
sequential acquisitions, position is less important---ALM takes care of
that---but the final results are still sensitive to the choice of the first
$n_{\mathrm{init}} = 8$ points, even though MLEs $\hat{\theta}$ are
recalculated after each selection.  Seeding the sequential design, which is
often glossed over as an implementation detail, can be crucial to good
performance in active learning.

Consequently, betadist, lhsbeta and random vastly outperform LHS and maximin.
The trouble with these space-filling seed designs is evident in the 90\%
quantile, which fails to improve even after many new design sites are added.
Too much spread in the initial design results in large $\hat{\theta}$s,  which
is reinforced by subsequent ALM acquisitions at the boundaries of the input
space.  The early behavior of maximin is particularly strange: getting worse
before better even in cases where sequential acquisitions lead to decent
results. It's 90\% quantile is eventually no worse than LHS's---quite poor.
The fact that maximin's average RMSPE is nearly as bad suggests that maximin
rarely recovers from that poor initial design.

\subsection{Expected improvement for optimization}
\label{sec:ei}

Here we show that betadist and lhsbeta initial designs are also superior in a
BO context similar to that used to find the best $\hat{\alpha}$ and
$\hat{\beta}$ settings in Section \ref{sec:betaopt}.  Specifically,
acquisitions are gathered via EI (\ref{eq:ei}) using a random five-start
scheme including the location of the best input setting (corresponding to
$\mu_{\min}$) from the previous iteration.  As a test function, we use the
so-called Greiwank function
\[
f_d(\mathbf{x}) = \sum_{i=1}^d\frac{x_i^2}{4000} 
- \prod_{i=1}^d \cos\left(\frac{x_i}{\sqrt{i}}\right)+1.
\]
For visualizations and further details, including {\sf R} implementation, see
the Virtual Library of Simulation Experiments: \url{https://www.sfu.ca/~ssurjano/griewank.html}.

A nice feature of the Greiwank is that it is defined for arbitrary input
dimension $d$, and is flexible about the bounds $b$ of the inputs, $\mathbf{x}
\in [-b,b]^d$.  These two settings, $b$ and $d$, together determine the
complexity of the response surface.  The global minima is always at the
origin, however the number of local minima grows quickly as $b$ and $d$ are
increased.  We utilize these knobs to vary the complexity of the function, in
order to span a range of optimization problems.  By varying the bounds $b$ in
particular, we vary the magnitude of the best lengthscale for the purpose of
surrogate modeling, and thereby create a situation where an initial design is
key to obtaining good performance in BO.

Our experimental setup is as follows.  We consider three
$(n_{\mathrm{init}},d)$-pairs from Figure \ref{f:betaopt} and track the
progress of EI-based BO measured by the lowest value of the objective found
over the sequential design iterations.  In each of one thousand MC
repetitions, we create initial $n_\mathrm{init}$-sized designs via maximin,
random, LHS, betadist and lhsbeta, with subsequent acquisitions handled by EI.
In accordance with the theory for convergence of EI-based BO
\citep{bull:2011}, we do not update $\hat{\theta}$ after each EI acquisition,
but fix it at the setting obtained immediately after the initial design.  This
has the benefit of accentuating the effect of the initial design, which suits
our illustrative purposes.  It is also more computationally efficient, leading
to an $O(n^3)$ calculation rather than $O(n^4)$ if MLEs are recalculated
regularly.  However the results are not much different under that latter
alternative.

To vary the complexity of the underlying optimization problem, and thus best
effective lengthscale for GP surrogate, we draw $b \sim \mathrm{Unif}(0, 10)$
at the start of each MC repetition.  In so doing, each of 1000 MC repetitions
targets a Greiwank function having a different degree of waviness, and number
of local optima. By holding $b$ fixed for each of the five initial design
choices, and subsequent EI-optimizations,  we create a setting wherein
pairwise $t$-tests can be used to adjudicate between those comparators.
Finally, all calculations were formed with methods built into the {\tt laGP}
package on CRAN.  Since we observe $f_d(\mathbf{x})$ without noise, no nugget
hyperparameters are required. Not presuming to know the randomly generated
scale $b$, we allow MLE calculations for $\hat{\theta}$ to search in a space
that would be appropriate for the largest settings, $\theta \in (0,
10\sqrt{d}]$, regardless of $b$.

\begin{table}[!ht]
\centering
\begin{tabular}{l|rrrrr}
  $n = 25$   & maximin & LHS & random & betadist & lhsbeta \\ \hline
  maximin & NA & 0.95  & 0.98 & $>$ 0.99 & $>$ 0.99  \\ 
  LHS & \textbf{0.048}  & NA & 0.67  & $>$ 0.99 & $>$ 0.99 \\ 
  random & \textbf{0.022} & 0.33 & NA & $>$ 0.99 & $>$ 0.99 \\ 
  betadist & \textbf{$<$ 1e-7} & \textbf{2e-5} & \textbf{8e-5} & NA & $>$ 0.99  \\ 
  lhsbeta & \textbf{$<$ 1e-7} & \textbf{$<$ 1e-7} & \textbf{$<$ 1e-7} & \textbf{$<$ 1e-7} & NA  \\ 
\end{tabular}\\

\vspace{0.25cm}

\begin{tabular}{l|rrrrr}
  $n=70$   & maximin & LHS & random & betadist & lhsbeta \\ \hline
 maximin & NA & $>$ 0.99 & $>$ 0.99 & $>$ 0.99 & $>$ 0.99  \\ 
 LHS & \textbf{5e-7} & NA & 0.89  & $>$ 0.99 & $>$ 0.99  \\ 
 random & \textbf{$<$ 1e-7} & 0.11 & NA & $>$ 0.99 & $>$ 0.99  \\ 
 betadist & \textbf{$<$ 1e-7} & \textbf{$<$ 1e-7} & \textbf{2e-6} & NA & $>$ 0.99  \\ 
 lhsbeta & \textbf{$<$ 1e-7} & \textbf{$<$ 1e-7} & \textbf{$<$ 1e-7} & \textbf{$<$ 1e-7} & NA  \\ 

\end{tabular}
\caption{Pairwise $t$-test $p$-value table for $(n_{\mathrm{init}} = 8, d=2)$ and two settings $n=25$ (top table) and $n=70$ (bottom).  Statistically significant $p$-values, i.e., below 5\%, are in bold.} 
\label{t:ein8d2}
\end{table}

Table \ref{t:ein8d2} summarizes results obtained from the
($n_{\mathrm{init}}=8$, $d=2$) case in two views: after $n=20$ total
acquisitions, and then after $n=70$.  The bolded $p$-values in the table(s)
are below the typical 5\% threshold.  Observe in both cases that random and
LHS design are consistently better than maximin, but betadist is significantly
better than all three.  Hybrid lhsbeta outperforms all of the others.
In other words, the story here is more or less the same as before. The only
substantial difference is that lhsbeta outperforms betadist.

\begin{table}[!ht]
\centering
\begin{tabular}{l|rrrrr}
  $n=50$    & maximin  & LHS & random & betadist &lhsbeta \\ \hline
  maximin & NA & $>$ 0.99 & $>$ 0.99 & $>$ 0.99 & $>$ 0.99 \\ 
  LHS & \textbf{$<$ 1e-7} & NA & 0.95 & $>$ 0.99 & $>$ 0.99  \\ 
  random & \textbf{$<$ 1e-7} & 5.3e-2 & NA & $>$ 0.99 & $>$ 0.99  \\ 
  betadist & \textbf{$<$ 1e-7} & \textbf{$<$ 1e-7} & \textbf{$<$ 1e-7} & NA & $>$ 0.99  \\ 
  lhsbeta & \textbf{$<$ 1e-7} & \textbf{$<$ 1e-7} & \textbf{$<$ 1e-7} &  \textbf{1e-3} & NA   \\ 
 
\end{tabular} \\

\vspace{0.25cm}

\begin{tabular}{l|rrrrr}
  $n=100$  & maximin  & LHS & random & betadist &lhsbeta \\ \hline
  maximin &NA  &  $>$ 0.99 &  $>$ 0.99 &  $>$ 0.99 & $>$ 0.99  \\ 
  LHS & \textbf{$<$1e-7} & NA &  $>$ 0.99 &  $>$ 0.99 &  $>$ 0.99  \\ 
  random & \textbf{$<$ 1e-7} & \textbf{3e-3} & NA &  $>$ 0.99 &  $>$ 0.99  \\ 
  betadist & \textbf{$<$ 1e-7} & \textbf{$<$ 1e-7} & \textbf{$<$ 1e-7} & NA &  $>$ 0.99  \\ 
  lhsbeta & \textbf{$<$ 1e-7} & \textbf{$<$ 1e-7} & \textbf{$<$ 1e-7} & \textbf{8e-3} & NA \\ 

\end{tabular}
\caption{Pairwise $t$-test $p$-value table for $(n_{\mathrm{init}} = 16, d=3)$ and two settings $n=50$ (top table) and $n=100$ (bottom).  Statistically significant $p$-values, i.e., below 5\%, are in bold.}
\label{t:ein16d3}
\end{table}

Table \ref{t:ein16d3} summarizes results from the $(n_{\mathrm{init}}=16,
d=3)$ case.  In higher dimension, the problem is more challenging with many
more local minima.  Both a bigger initial design, and a larger run of EI
acquisitions is necessary in order to obtain reliable results.  At $n=50$ the
pecking order is similar: maximin, LHS, betadist, lhsbeta---all statistically
significant at the 5\% level. Random outperforms LHS, but not significantly so
at the 5\% level.

\begin{table}[!ht]
\centering
\begin{tabular}{l|rrrrr}
  $n=200$     & maximin  & LHS & random & betadist &lhsbeta\\ \hline
  maximin & NA & $>$ 0.99 & $>$ 0.99 & $>$ 0.99 & $>$ 0.99 \\ 
  LHS & \textbf{$<$ 1e-7} & NA & 0.43 & $>$ 0.99 & $>$ 0.99 \\ 
  random & \textbf{$<$ 1e-7} & 0.57 & NA & $>$ 0.99 & $>$ 0.99  \\ 
  betadist & \textbf{$<$ 1e-7} & \textbf{2e-4} & \textbf{2e-4} & NA & $>$ 0.99  \\ 
  lhsbeta & \textbf{$<$ 1e-7} & \textbf{$<$ 1e-7} & \textbf{$<$ 1e-7} & \textbf{$< $1e-7} & NA  \\  
\end{tabular} \\

\vspace{0.25cm}

\begin{tabular}{l|rrrrr}
  $n=500$    & maximin  & LHS & random & betadist &lhsbeta \\ \hline
  maximin & NA & $>$ 0.99 & $>$ 0.99 & $>$ 0.99 & $>$ 0.99  \\ 
  LHS & \textbf{$<$1e-7} & NA & 0.25 & $>$ 0.99 & $>$ 0.99 \\ 
  random & \textbf{$<$ 1e-7} & 0.75  & NA & $>$ 0.99 & $>$ 0.99  \\ 
  betadist & \textbf{$<$ 1e-7} & \textbf{8e-3} & \textbf{2e-3} & NA & $>$ 0.99  \\ 
  lhsbeta & \textbf{$<$ 1e-7} & \textbf{$<$ 1e-7} & \textbf{$<$ 1e-7} & \textbf{$<$ 1e-7} & NA  \\ 
\end{tabular}
\caption{Pairwise $t$-test $p$-value table for $(n_{\mathrm{init}} = 32, d=4)$ and two settings $n=200$ (top table) and $n=500$ (bottom).  Statistically significant $p$-values, i.e., below 5\%, are in bold.}
\label{t:ein32d4}
\end{table}

Finally, Table \ref{t:ein32d4} summarizes the $(n_{\mathrm{init}}=32, d=4)$
case with $n=200$ and $n=500$.  Except when the randomly chosen $b$ is very
small, this setting represents an extremely difficult optimization with dozens
of local minima.  A large number of samples is required to obtain decent
global BO results.  The story here is very similar to Tables
\ref{t:ein8d2}--\ref{t:ein16d3}.

\section{Discussion}
\label{sec:discuss}

We have described a new scheme for design for surrogate modeling of
computer experiments based on pairwise distance distributions.  The idea was
borne out of the occasionally puzzling behavior of more conventional maximin
and LHS designs, especially as deployed as initial designs in a sequential
setting.  Maximin designs, and to a certain extent LHS, lead to a highly
irregular pairwise distance distribution which all but precludes the
estimation of small lengthcales except when the design is very large.  By
deliberately targeting a simpler family of unimodal distance distributions we
have found that it is possible to avoid that puzzling behavior, obtain a more
accurate estimate of the lengthscale, and ultimately make better predictions and
sequential design decisions.  For reproducibility, the code behind our
empirical work is provided in an open Git repository on Bitbucket:
\url{https://bitbucket.org/gramacylab/betadist}.

We proposed an optimization strategy for finding the best distance
distributions within the Beta family conditional on the design setting,
specified kernel family, design size ($n$) and input dimension ($d$).  Many
potential avenues for further investigation naturally suggest themselves.  For
simplicity, we limited our study to the isotropic Gaussian family.  One could
check that similar results hold for other common families like the Mat\'ern. A
more ambitious extension would be to separable structures where there is a
lengthscale for each input coordinate: $\theta_1, \dots, \theta_d$.  Obtaining
appropriate pairwise distance distributions in each coordinate simultaneously
could prove difficult, especially in small-$n$ large-$d$ settings.
\blu{However, we speculate that the problem could be effectively reduced down to
$d$ univariate ones.}  Considering nugget hyperparameters in the optimization
would add yet another layer of complication.  In that setting, we may wish to
consider replication (i.e., zero-inflated distance distributions) as a means
of separating signal from noise \citep{hetGP2}.

\blu{Many response surfaces from simulations of industrial systems are
exceedingly smooth and slowly varying over the study region of interest. Such
knowledge, when available, could translate into an {\em a priori} belief about
large lengthscales $\theta$, or even a lower bound on $\theta$.  In our
empirical work, and searches for optimal $\mathrm{betadist}_{n,d}(\alpha,
\beta)$ through simulated $\theta$-values, we took a lower bound on $\theta$
of effectively zero.  However, we see no reason why a different lower bound
couldn't be applied.  We speculate that narrowing the range of $\theta$,
especially toward the upper end, would result in an organic preference for
larger pairwise distances through the search for optimal $(\hat{\alpha},
\hat{\beta})$, and that these designs will perform more similarly to
space-filling ones like maximin.}

Another family of target distance distributions, i.e., besides the Beta, could
prove easier to optimize over, or otherwise lead to better designs.  A
higher-powered search for designs, besides random swapping, might mitigate the
computational burden of finding optimal distance-distributed designs which
becomes problematic when $n$ is large.  Some researchers have recently had
success with particle swarm optimization (PSO) in design settings, like
minimax design \citep{chen:etal:2015}, which might port well to the
distance-distribution setting and the lhsbeta hybrid.

Perhaps the most important take home message from this manuscript is that
maximin designs can be awful.  LHSs are better, because they avoid a
multi-modal distance distribution and, simultaneously, a degree of aliasing
through their one-dimensional uniformity property.  However, we argue that the
most important thing is to have a good design for hyperparameter inference,
which neither method targets directly.  In fact, random design is better than
both in this respect, which is perhaps surprising.  If you assume to know the
hyperparameters, then LHS and maximin are great.  It's worth noting that
ascribing physical or interpretive meaning to lengthscale hyperparameters can
be extremely challenging. Therefore, it is hard to imagine that one could
consistently choose appropriate lengthscales without help from automatic
procedures like MLE---which, of course, need a design.

\subsubsection*{Acknowledgments} 

Authors BZ, DAC and RBG gratefully acknowledge funding from a DOE LAB 17-1697
via subaward from Argonne National Laboratory for SciDAC/DOE Office of Science
ASCR and High Energy Physics.  RBG and DAC recognize partial support from
National Science Foundation (NSF) grants DMS-1849794 and DMS-1821258.  RBG
acknowledges partial support from NSF DMS-1621746.

\bibliographystyle{jasa.bst}
\bibliography{unifdist}

\end{document}